\makeatletter
\declare@file@substitution{revtex4-1.cls}{revtex4-2.cls}
\makeatother

\PassOptionsToPackage{dvipsnames}{xcolor}
\documentclass[twocolumn]{aastex631}

\usepackage{soul}
\usepackage{xcolor}
\usepackage{xspace,graphicx,url}
\usepackage{natbib}

\usepackage{bold-extra}
\usepackage{hyperref}
\definecolor{hlblue}{RGB}{89,255,255} \definecolor{hlyellow}{RGB}{253,255,50} \definecolor{rlink}{rgb}{1,0,0} \definecolor{glink}{rgb}{0,0.9,0} \definecolor{blink}{rgb}{0,0,1} \definecolor{mlink}{rgb}{1,0,1} \hypersetup{linkcolor=blink,citecolor=blink,filecolor=blink,urlcolor=blink}
\hypersetup{linkcolor=rlink,citecolor=glink,filecolor=mlink,urlcolor=blink}

\usepackage{amsmath}
\usepackage{tabularx}
\usepackage{booktabs}
\usepackage{threeparttable}
\usepackage{kantlipsum}
\usepackage{enumitem}

\graphicspath{{./}}

\makeatletter

\newcommand{\targf}{Swift J1727.8$-$1613\xspace}
\newcommand{\targ}{J1727\xspace}
\newcommand{\pcalf}{J1939$-$6342\xspace}

\newcommand{\scalf}{PKS J1733$-$1304\xspace}
\newcommand{\scal}{J1733\xspace}
\newcommand{\HI}{H\,{\sc i}\xspace}

\newcommand{\dnearstd}{3.6 \pm 0.3}

\newcommand{\dtanstd}{8.20 \pm 0.03}

\newcommand{\vtanstd}{93.9 \pm 3.3}

\newcommand{\dnearunc}{3.6 \pm 0.3 \, ({stat}) \pm 2.3 \, ({sys})}
\newcommand{\EBmVNH}{0.27\,\pm\,0.01}
\newcommand{\EBmVNUV}{0.37 \pm 0.01 \, ({stat}) \pm 0.025 \, ({sys})}
\newcommand{\EBmVuplim}{0.4}

\newcommand{\dfinal}{5.5^{+1.4}_{-1.1}}
\newcommand{\drange}{$4.4\,\text{--}\,6.9$\,kpc\xspace}
\newcommand{\Porb}{P_{\rm orb}\xspace}
\newcommand{\kps}{km\,s$^{-1}$\xspace}
\newcommand{\angstrom}{\AA\xspace}
\newcommand{\EBmV}{E(B\!-\!V)}
\newcommand{\NH}{N_{\rm H}}
\newcommand{\kickv}{190\,\pm\,30}

\defcitealias{MataSanchez_etal_2024_Swift_J1727_Inflows_Outflows}{MS24}
\newcommand{\mata}{{\citetalias{MataSanchez_etal_2024_Swift_J1727_Inflows_Outflows}}\xspace}
\defcitealias{MataSanchez_etal_2025_Swift_J1727_Dynamically_Confirmed_BH}{MS25}
\newcommand{\matat}{{\citetalias{MataSanchez_etal_2025_Swift_J1727_Dynamically_Confirmed_BH}}\xspace}

\usepackage[none]{hyphenat}
\usepackage[normalem]{ulem}

\makeatother

\begin{document}

\title{On the distance to the black hole X-ray binary \targf}

\author[0009-0004-0093-0096]{Benjamin J. Burridge}
\affiliation{International Centre for Radio Astronomy Research, Curtin University, GPO Box U1987, Perth WA 6845, Australia}

\author[0000-0003-3124-2814]{James C. A. Miller-Jones}
\affiliation{International Centre for Radio Astronomy Research, Curtin University, GPO Box U1987, Perth WA 6845, Australia}

\author[0000-0003-2506-6041]{Arash Bahramian}
\affiliation{International Centre for Radio Astronomy Research, Curtin University, GPO Box U1987, Perth WA 6845, Australia}

\author[0000-0003-3165-6785]{Steve R. Prabu}
\affiliation{International Centre for Radio Astronomy Research, Curtin University, GPO Box U1987, Perth WA 6845, Australia}
\affiliation{Astrophysics, Department of Physics, University of Oxford, Keble Road, Oxford, OX1 3RH, UK}

\author{Reagan Streeter}
\affiliation{International Centre for Radio Astronomy Research, Curtin University, GPO Box U1987, Perth WA 6845, Australia}

\author[0000-0002-5870-0443]{Noel Castro Segura}
\affiliation{Department of Physics, University of Warwick, Gibbet Hill Road, Coventry CV4 7AL, UK}

\author[0000-0003-1038-9104]{Jes\'us M. Corral Santana}
\affiliation{European Southern Observatory, Alonso de C\'ordova 3107, Casilla 19001 Vitacura, Santiago, Chile}

\author[0000-0002-1116-2553]{Christian Knigge}
\affiliation{School of Physics \& Astronomy, University of Southampton, Southampton SO17 1BJ, UK}

\author[0000-0002-0333-2452]{Andrzej Zdziarski}
\affiliation{Nicolaus Copernicus Astronomical Center, Polish Academy of Sciences, Bartycka 18, PL-00-716 Warszawa, Poland}

\author[0000-0003-0245-9424]{Daniel Mata S\'anchez}
\affiliation{Instituto de Astrof\'isica de Canarias, E-38206 La Laguna, Tenerife, Spain}
\affiliation{Departamento de Astrof\'isica, Univ. de La Laguna, E-38206 La Laguna, Tenerife, Spain}

\author[0000-0002-4039-6703]{Evangelia Tremou}
\affiliation{National Radio Astronomy Observatory, Socorro, NM 87801, USA}

\author[0000-0002-0426-3276]{Francesco Carotenuto}
\affiliation{INAF-Osservatorio Astronomico di Roma, Via Frascati 33, I-00076, Monte Porzio Catone (RM), Italy}

\author[0000-0002-5654-2744]{Rob Fender}
\affiliation{Astrophysics, Department of Physics, University of Oxford, Keble Road, Oxford, OX1 3RH, UK}

\author[0000-0002-5319-6620]{Payaswini Saikia}
\affiliation{Center for Astrophysics and Space Science, New York University Abu Dhabi, PO Box 129188, Abu Dhabi, UAE}

\correspondingauthor{Benjamin J. Burridge}
\email{benjamin.burridge@icrar.org}

\begin{abstract}

We review the existing distance estimates to the black hole X-ray binary \object{\targf}, present new radio and near-UV spectra to update the distance constraints, and discuss the accuracies and caveats of the associated methodologies.

We use line-of-sight \HI absorption spectra captured using the MeerKAT radio telescope to estimate a maximum radial velocity with respect to the local standard of rest of $24.8 \pm 2.8$\,\kps for \targf, which is significantly lower than that of a nearby extragalactic reference source.
From this we derive a near kinematic distance of $d_{\rm{near}} = \dnearunc$\,kpc as a lower bound after accounting for additional uncertainties given its Galactic longitude and latitude, $(l, b) \approx (8.6^{\circ}, 10.3^{\circ})$. 

Near-UV spectra from the Hubble Space Telescope's Space Telescope Imaging Spectrograph allows us to constrain the line-of-sight colour excess to $\EBmV = \EBmVNUV$.
We then implement this in Monte Carlo simulations and present a distance to \targf of $\dfinal$\,kpc, under the assumption that the donor star is an unevolved, main sequence K4($\pm$1)V star.
This distance implies a natal kick velocity of $\kickv$\,\kps and therefore an asymmetrical supernova explosion within the Galactic disk as the expected birth mechanism.

A lower distance is implied if the donor star has instead lost significant mass during the binary evolution.
Hence, more accurate measurements of the binary inclination angle or donor star rotational broadening from future observations would help to better constrain the distance.

\end{abstract}


\section{Introduction} \label{sec:intro}

Distance is an important parameter in the study of all astrophysical objects.
For Galactic low-mass X-ray binaries (XRBs), accurate distances allow for better estimation of other parameters, such as the peak Eddington luminosity ($L_{\rm{Edd}}$) fraction (ELF) and jet parameters including physical size scales, inclination angles, and speeds.

Reliably measuring the distance to newly discovered XRBs can be challenging.
For instance, across the different phases of an outburst, the system's luminosity can fluctuate, often in and out of detectable levels.
This inconsistent detectability can preclude the measurements required for accurate distance determination.

\subsection{Distance methods} \label{sec:intro_distance}

Distance determination techniques applicable to XRBs can be broadly categorised into three groups: (1) astrometric or kinematic methods; (2) approaches that use observations of the donor star, X-ray source, or jets; or (3) techniques based on propagation effects. We summarise these in Sections \ref{sec:intro_astrometrykinematics}--\ref{sec:intro_propagation}. We then discuss in more detail the two techniques that we employ that exploit the relationships between the distance and \HI absorption in radio observations (Section \ref{sec:intro_HI}), and the colour excess or reddening, $\EBmV$, as measured using near-UV observations (Section \ref{sec:intro_EBmV}).

\subsubsection{Astrometry and kinematics} \label{sec:intro_astrometrykinematics}

High-significance XRB parallax measurements with {\em Gaia} \citep{Gaia_Collaboration_etal_2016_The_Gaia_Mission, Gandhi_etal_2019_Gaia_DR2_distances_peculiar_velocities_for_GBHTs, Atri_etal_2019_BH_XRB_natal_kick_velocity_distribution, Arnason_etal_2021_Distances_Galactic_XRBs_Gaia_DR2} or very long baseline interferometry (VLBI) at radio wavelengths \citep[e.g.;][]{Miller-Jones_etal_2009_V404_Cyg_parallax, Reid_etal_2011_Cyg_X-1_Trig_VLBI_parallax, Reid_etal_2014_GRS_1915_Parallax_Distance_Revised_BH_Mass, Atri_etal_2020_MAXI_J1820_parallax, Miller-Jones_etal_2021_Cyg_X-1_parallax_revised, Reid+Miller-Jones_2023_Cyg_X-3_GRS_1915_distances_parallax} are the gold standard for measuring distances to Galactic XRBs.
However, radio parallaxes can be impeded by line-of-sight scatter broadening for XRBs located in the Galactic Plane (GP).
In fact, extinction in the GP and the faintness of quiescent XRBs precludes {\em Gaia} distances in many cases.
Given the typical kiloparsec (kpc) distances of XRBs, sub-milliarcsecond precision is required \citep{Tetarenko_etal_2016_WATCHDOG}.
These methods also require observations that span an extended timeframe, which is not always possible for XRB outbursts.

Alternatively, kinematic distance methodologies use measured proper motions and velocities predicted using the Galactic rotation model to infer the most likely distance.
Despite relying on the assumption of low peculiar velocities relative to the local standard of rest (LSR), these can provide reliable distances in some circumstances \citep{Reid_2022_On_the_accuracy_of_3D_kinematic_distances} and can be useful for XRBs \citep[e.g.;][]{Dhawan_etal_2007_GRS_1915_Kinematics_BH_XRB,Reid+Miller-Jones_2023_Cyg_X-3_GRS_1915_distances_parallax}.

\subsubsection{Stellar, X-ray, and jet observations} \label{sec:intro_stellarxrays}

One can use optical spectroscopy of the XRB donor star to estimate the distance \citep[e.g.;][]{Dubus_etal_2001_XTE_J1118+480_Optical_Spectroscopy_Outburst, Jonker+Nelemans_2004_Distances_Galactic_LMXB_spectroscopy,Charles_etal_2019_Swift_J1357.2-0933_HeII_outflows_2017_outburst}.
With measured values for the donor star's absolute and apparent magnitudes and the extinction along the line of sight, one can use the distance modulus \citep[e.g.;][]{MataSanchez_etal_2024_Swift_J1727_Inflows_Outflows, MataSanchez_etal_2025_Swift_J1727_Dynamically_Confirmed_BH} to infer the distance, via
\begin{equation} \label{eqn:distance_modulus}
    d = 10^{(m - M - A + 5)/5}
\end{equation}
where $d$ is the distance, $m$ is the apparent magnitude, $M$ is the absolute magnitude, and $A$ is the extinction.

Obscuration of optical light is pronounced for targets residing in the GP due to increased interstellar dust, making donor stars difficult to identify, and introducing additional uncertainty to the distance modulus equation.
However, for targets outside the GP, extinction can be harder to estimate accurately.

X-ray luminosities of XRB outbursts during soft-to-hard and hard-to-intermediate state transitions have been observed to occur at somewhat consistent ELFs, albeit with factor of $\sim \! 3$ scatter in these measurements \citep{Kalemci_etal_2013_BHT_distance_state_transitions,Tetarenko_etal_2016_WATCHDOG, VahdatMotlagh+Kalemci+Maccarone_2019_State_transition_luminosities_Galactic_BHT_outburst_decay}.
X-ray studies during these transitions allow one to compare the measured and expected intrinsic luminosities and thereby estimate the distance \citep[e.g.;][]{Abdulghani_etal_2024_Swift_J1727_X-ray_Transition_Luminosities_ELF}.

Further X-ray methods exist, including the combination of X-ray spectroscopy and the distance dependency of accretion disk spectral fits, which \citet{Hynes_etal_2002_XTE_J1859_Accretion_Disc_X-ray_Distance} applied to constrain the distance to XTE J1859+226.
Additionally, \citet{Powell_etal_2007_Mass_Transfer_LMX_Transient_Decays_Distance} used the time-scale of outburst decay X-ray light curves to estimate the absolute luminosity at a characteristic time and therefore provide a measure of the distance.

In the radio band, the proper motions of two-sided jets can be combined to place an upper limit on the source distance \citep[e.g.,][]{Mirabel+Rodriguez_1994_GRS_1915_Nature_Superluminal}.

\subsubsection{Propagation and the Interstellar Medium} \label{sec:intro_propagation}

X-rays produced by XRB flares will propagate outwards and may subsequently scatter off intervening interstellar dust clouds.
Provided the distances to these dust clouds can be determined, one can combine this information with analysis of the time delays and intensities of these expanding X-ray dust scattering rings to measure the distance to the source \citep[e.g.;][]{Heinz_etal_2015_Cir_X-1_LotR_Kinematic_Distance_X-ray_Light_Echo, Beardmore_etal_2016_LotR_RotK_X-ray_dust_scattering_rings_V404_Cyg, Lamer_etal_2021_X-ray_scattering_distance_MAXI_J1348}.

X-ray absorption features in observed spectra can be used to measure the hydrogen column density, $\NH$.
When coupled with hydrogen distribution models one can infer the distance to the source.

The relation between $\EBmV$ and the aforementioned extinction along the line of sight can be used to inform distance modulus calculations \citep[e.g.;][]{Schlafly_Finkbeiner_2011_EBmV_with_SDSS}.
The inverse relation between $\EBmV$ and the distance therefore allows constraints on one to constrain the other.
We examine and implement this method in Sections \ref{sec:intro_EBmV} and \ref{sec:discussion_EBmVdist} respectively.

Lastly, line-of-sight \HI absorption has long been used as an XRB distance estimator 
\citep[e.g.;][]{Dickey_1983_Cygnus_X-3_HI_distance, Lockman_etal_2007_SS433_HI_distance_ISM, Chauhan_etal_2019_MAXI_J1535-571_HI, Chauhan_etal_2021_MAXI_J1348-630_HI}.
We explore this method further in Section \ref{sec:intro_HI} below and apply it in Sections \ref{sec:data_radio}, \ref{sec:results_radio}, and \ref{sec:discussion_kd}.

\subsubsection{H\,{\footnotesize I} absorption} \label{sec:intro_HI}

Our first distance method uses \HI absorption, which can be observed when clouds of neutral hydrogen along the line of sight absorb the broadband continuum emission produced by the target at the \HI frequency in their rest frame.
These clouds move with different velocities relative to us along the line of sight, due to the rotation of the Milky Way, as well as other effects such as non-circular streaming motions that we assume to be minimal.
The more clouds that are intersected by the line of sight, the more \HI absorption features that are imprinted at different frequencies on the observed radio spectrum.
One benefit of \HI absorption over parallax is that the required data can be gathered within a much shorter timeframe; a single observation can suffice should the observed source be particularly bright.

The Doppler-shifted frequencies can be converted into LSR velocities and compared to the Milky Way rotation curve.
The maximum velocity occurs at the tangent point, where the rotational velocity is entirely along the line of sight.
Identical velocities are seen either side of this maximum, giving rise to an ambiguity in mapping observed absorption velocities to distances within the solar circle.
A maximum observed velocity that is less than the tangent point velocity could correspond to a near kinematic distance before the tangent point, or a far kinematic distance beyond the tangent point \citep[e.g.;][Figure 4]{Wenger_etal_2018_Kinematic_Distances_MC}.

To resolve this kinematic distance ambiguity, one must observe the target but also at least one extragalactic reference source close enough to the target in the sky such that any differences in the anticipated \HI distributions along the lines of sight are minimised.
The emission from the reference source will have passed through all Galactic \HI clouds along the line of sight, with clouds outside the solar circle imprinting absorption velocities of the opposite sign.
Any absorption present in the reference spectrum but absent in the target spectrum then allows us to place an upper limit on the distance.

\subsubsection{$\EBmV$} \label{sec:intro_EBmV}

Our second distance method relies on the reddening caused by interstellar dust preferentially scattering shorter wavelengths.
This can be determined by subtracting the observed difference between blue and visible magnitudes, $B$ and $V$, to quantify the reddening along the line of sight.

$\EBmV$ can be calculated using various relations between it and interstellar absorption lines \citep[e.g.;][]{Munari_Zwitter_1997_EWs_NaI_KI_EBmV, Wallerstein_Sandstrom_Gredel_2007_8621_Angstroms_EBmV}, or $\NH$ \citep[e.g.;][]{MataSanchez_etal_2025_Swift_J1727_Dynamically_Confirmed_BH}.
It can be estimated from Galactic dust maps, both 2-dimensional \citep[2D; e.g.;][]{Schlegel_Finkbeiner_David_1998_SFD_dust_map_via_IR_emission_for_reddening_and_CMBRFs, Chiang_2023_Corrected_SFD_CSFD_dust_map_minimal_extragalactic_contamination} and 3-dimensional \citep[3D; e.g.;][]{Green_etal_2019_3D_dust_map_Gaia_Pan-STARRS1_2MASS, Edenhofer_etal_2024_pc_scale_3D_dust_map}.
Near-UV spectra can also be used as implemented in Sections \ref{sec:data_NUV} and \ref{sec:results_NUV}.

With a value for $\EBmV$ one can derive the extinction.
For example, it is common to use $R_V = 3.1$ as a Galactic average with
\begin{equation}
    \label{eqn:EBmV_to_AV}
    A_V = R_V \, \EBmV
\end{equation}
to convert reddening to the total extinction along the line of sight or vice versa \citep{Savage_Mathis_1979_Properties_Interstellar_Dust,Fitzpatrick_2004_Interstellar_Extinction_Milky_Way}.
When combined with measurements of absolute and apparent magnitudes in Equation \ref{eqn:distance_modulus}, the distance can then be calculated.

\subsection{\targf} \label{sec:intro_targ}

\targf (\targ), located at $(l, b) = (8.641502^{\circ}, 10.254899^{\circ})$, was first detected as an X-ray transient on 24 August 2023 \citep{Negoro_etal_2023_ATel_16205_Swift_J1727_MAXI/GSC_detection}.
Bright radio emission was observed within a couple of days \citep{Miller-Jones_etal_2023_ATel_16211_Swift_J1727_VLA_radio_detection}, which continued to brighten through early September \citep{Bright_etal_2023_ATel_16228_Swift_J1727_ATA_detection}.
Analysis of observations in late August and early September revealed a bright core and a large two-sided, asymmetrical jet \citep{Wood_etal_2024_Swift_J1727_Largest_Resolved_Jet_XRB}.
Radio monitoring in early October suggested radio quenching and subsequent flaring \citep{Miller-Jones_etal_2023_ATel_16271_Swift_J1727_Radio_quenching_subsequent_flaring}.

This event was deemed a low-mass XRB outburst \citep{Castro-Tirado_etal_2023_ATel_16208_Swift_J1727_Optical_spectroscopy_confirms_BH-LMXB}, and its high radio brightness made \targ a suitable target for \HI absorption measurements.
Since then, the compact object has been dynamically confirmed to be a black hole (BH) \citep[][MS25 hereafter]{MataSanchez_etal_2025_Swift_J1727_Dynamically_Confirmed_BH}.

Further studies of the outburst revealed that it produced relativistic jets that are the largest resolved jets in an XRB to date \citep{Wood_etal_2024_Swift_J1727_Largest_Resolved_Jet_XRB}.
The ejection of transient jets in \targ has also been shown to have occurred simultaneously with a bright X-ray flare and a sudden change in the X-ray properties of the accretion inflow \citep{Wood_etal_2025_Swift_J1727_Ejection_Transient_Jets_Time_Dependent_Visibility_Modelling}.

The {\em Gaia} optical counterpart for \targf currently has proper motion but no parallax.
A VLBI radio parallax will not be possible with the observations taken to date, as the XRB has already returned to the quiescent state.

\subsubsection{Current distance estimates} \label{sec:intro_currentstate}

\citet{Abdulghani_etal_2024_Swift_J1727_X-ray_Transition_Luminosities_ELF} estimated a distance of $1.52^{+0.85}_{-0.61}$\,kpc from a Bayesian approach of soft-state X-ray modelling.
This appeared to align with \citet{Veledina_etal_2023_Swift_J1727_X-ray_pol_flux_scaling_distance} who used X-ray flux scaling arguments to provide an early estimate of approximately $1.5$\,kpc.
However, \citet{Abdulghani_etal_2024_Swift_J1727_X-ray_Transition_Luminosities_ELF} concede that their distance estimate may be underestimated by up to $\sim \! 70\%$, given that only soft-state data and no state-transition information was used.

\citet[][MS24 hereafter] {MataSanchez_etal_2024_Swift_J1727_Inflows_Outflows} used donor star magnitudes in conjunction with various relations in the literature to derive values for the parameters in Equation \ref{eqn:distance_modulus}.
These included the relation between the interstellar Ca\,{\sc ii} doublet (H and K) and the distance to early-type stars per \citet{Megier_etal_2009_Interstellar_CaII_and_Distance}.
This was calibrated using objects within a few hundred pc from the GP (up to 450\,pc), slightly below but still consistent with their final inferred height for \targ.
The authors also used the relations between the equivalent widths of the interstellar line K\,{\sc i} 7699\,\angstrom and diffuse interstellar band at 8621 \angstrom and $\EBmV$ per \citet{Munari_Zwitter_1997_EWs_NaI_KI_EBmV} and \citet{Wallerstein_Sandstrom_Gredel_2007_8621_Angstroms_EBmV} respectively.
However, these relations resulted in particularly large values for $\EBmV$ at $0.8 \pm 0.3$ and $0.9 \pm 0.3$ respectively.
Lastly, the relation between hydrogen column density $\NH$ and $V$-band extinction $A_V$ per \citet{Guver_etal_2009_Optical_extinction_hydrogen_column_density} was combined with Equation \ref{eqn:EBmV_to_AV} to infer $\EBmV = 0.47 \pm 0.13$. 
These values of $\EBmV$ cover a wide range and some have large associated uncertainties, which propagate to the distance constraints.
We discuss this further in Section \ref{sec:discussion_EBmVvalidation} using our near-UV results and Galactic dust maps.

Following the above, \mata calculated the weighted mean of the resulting distances to be $d = 2.7 \pm 0.3$\,kpc.
\matat then directly measured the orbital period, $\Porb$, and reported the best fitting spectral type template of K4($\pm$1)V for a donor star that is partially veiled by the accretion disk.
Using this, they revised the absolute $r$-band magnitude to $M_r = 6.6\pm 0.5$ mag.
They also measured an apparent $r$-band magnitude of $m_r = 21.13 \pm 0.05$ mag, and presented an updated consolidated weighted mean distance of $d = 3.4 \pm 0.3$\,kpc.

\subsubsection{Contents} \label{sec:intro_contents}

\begin{table*}
    \hspace{-1.5cm}
    \begin{tabular}{lcccccccc}
        \toprule
        MJD & Observation & Observation & Exposure & L-band & Centre & Total & Channel & \targ peak \\
        & start date & start time & time & mode & frequency & bandwidth & width & flux density \\
        & (dd-mm-yyyy) & (hh:mm:ss) & (mm:ss) & & (MHz) & (MHz) & (kHz) & (mJy) \\
        \midrule
        60183 & 27-08-2023 & 15:27:59.6 & 14:55.6 & Standard & 1283.9869 & 856 & 26.123 & $49.7 \pm 0.2$ \\
        60193 & 06-09-2023 & 15:06:32.1 & 14:56.9 & Zoom & 1419.9984 & 107 & 3.265 & $97.5 \pm 0.4$ \\
        60231 & 14-10-2023 & 12:40:20.2 & 14:55.6 & Standard & 1283.9869 & 856 & 26.123 & $836.6 \pm 2.3$ \\
        60233 & 16-10-2023 & 15:50:13.8 & 14:56.9 & Zoom & 1419.9984 & 107 & 3.265 & $104.3 \pm 0.4$ \\
        \bottomrule
    \end{tabular}
    \caption{
        A summary of MeerKAT observation parameters. All times are in Coordinated Universal Time (UTC).
        The uncertainties on the peak flux densities are the root-mean-square (RMS) noise in the continuum image of the \targ field.
        }
    \label{tab:observations}
\end{table*}

In Section \ref{sec:data} we detail the methods used and data obtained.
In Section \ref{sec:results} we present our \HI absorption and near-UV spectra.
In Section \ref{sec:discussion} we discuss the interpretation of our results in constraining the distance to \targ, along with various caveats and implications for natal kick velocities and Eddington luminosity fractions.
In Section \ref{sec:conclusions} we present our suggested distance to \targ.

\section{Observations and Data Reduction} \label{sec:data}

\subsection{MeerKAT radio data} \label{sec:data_radio}

We observed \targ as part of the \textbf{T}he \textbf{hun}t for \textbf{d}ynamic and \textbf{e}xplosive \textbf{r}adio transients with Meer\textbf{KAT}\footnote{\url{http://science.uct.ac.za/thunderkat}} \citep[ThunderKAT;][]{Fender_etal_2016_ThunderKAT} large survey project and its successor, X-KAT (PI Fender).

We conducted 1--2 GHz (L-band) radio observations of the \targ field using the South African Square Kilometre Array precursor radio telescope, MeerKAT \citep{Camilo_2018_Afican_star_joins_radio_astronomy_firmament_MeerKAT}, between 27 August and 16 October 2023.
Our measured flux densities for \targ were in the range 50--837 mJy due to radio flaring of the source.
Further details of these observations are provided in Table \ref{tab:observations}.

All observations were performed with the L-band receiver; two using MeerKAT's standard ``32k'' mode, and two using the  ``32k zoom'' mode, hereafter referred to as ``32k-S'' and ``32k-Z'' respectively.
While each mode contains 32,768 channels, the 32k-Z mode has channel bandwidths that are eight times smaller and thus provides an eight-fold increase in frequency resolution.
We alternated our observation scans between \targ and the phase calibrator, \scalf (\scal hereafter), with a single scan of the bandpass and flux calibrator \pcalf in each observation.

Having obtained MeerKAT observations of \targ during which it was sufficiently bright (i.e., $\gtrsim \! 50$\,mJy) at 1.4 GHz, we compute \HI absorption spectra by processing the radio data to create a radio spectrum that includes the frequency of the \HI spectral line, $f_{\textsc{Hi}} = 1420.30575177$ MHz.
We convert Doppler shifts in frequency to line-of-sight LSR velocities.
We then compare the maximum positive or negative velocity observed in the resulting spectra with the Milky Way rotation curve to generate estimates of the kinematic distance via the source code for the Kinematic Distance Calculation Tool\footnote{\url{http://www.treywenger.com/kd/}}\textsuperscript{,}\footnote{\url{http://github.com/tvwenger/kd}} \citep[KDCT;][]{Wenger_2018_SW_KDCT_Zenodo_DOI_1166001}.

\subsubsection{Reference source selection} \label{sec:data_reference}

Due to the lack of bright ($\gtrsim \! 50$\,mJy) extragalactic background sources in the field i.e., within $1^{\circ}$ of \targ, we used the bright ($S_0 \approx 6$\,Jy) phase calibrator \scal to derive our reference \HI absorption spectra.
With \scal located at $(l,b) \approx (12.03^{\circ},10.81^{\circ})$, the two fields are only $3.4^{\circ}$ apart, primarily in Galactic longitude.
As \scal is extragalactic, observations allow us to probe the full set of \HI clouds along a nearby line of sight (see also Section \ref{sec:discussion_height} regarding \HI scale height).

\begin{figure*}[t]
    \centering
    \includegraphics[width=0.99\textwidth]{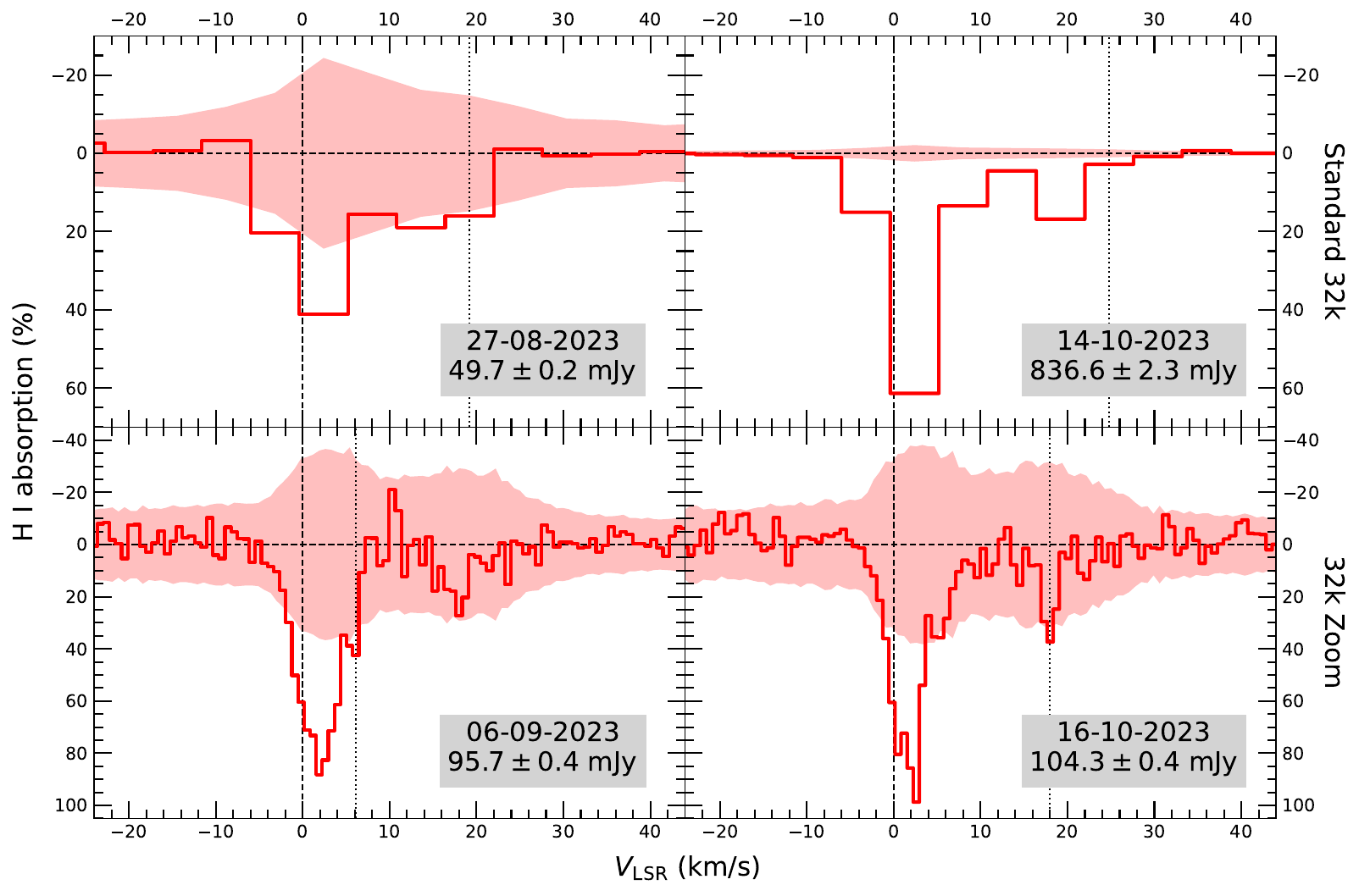}
    \caption{
        Our \HI absorption spectra along of the line of sight towards \targ.
        The top two plots are the 32k-S spectra, and the bottom two plots are the 32k-Z spectra, with LSR velocity bin widths that are 5.6 and 0.7 \kps wide respectively.
        The dashed lines indicate the origins of the $x$- and $y$-axes.
        The $y$-axis represents the residuals after subtracting the continuum emission from the data, with $3\sigma$ uncertainties indicated by the shaded areas, both of which are presented as a percentage of the continuum flux density of the source.
        The vertical dotted lines indicate the maximum velocity taken as the mid-point of the bin at which significant ($>\!3\sigma$) absorption is observed.
        In chronological order, these maximum velocities are
            $19.2 \pm 2.8$,
            $6.1 \pm 0.4$,
            $24.8 \pm 2.8$, and
            $18.0 \pm 0.4$\,\kps.
        The significantly greater SNR of the 14-10-2023 spectrum meant we were able to observe significant ($>\!3\sigma$) \HI absorption to greater velocities than observed in the 32k-Z spectra.
        }
    \label{fig:HI_Spectra_SwiftJ1727}
\end{figure*}

\subsubsection{Data reduction} \label{sec:data_reduction}

We undertook all data reduction on the Ilifu research cloud infrastructure managed by the Inter-University Institute for Data Intensive Astronomy (IDIA)\footnote{\url{http://idia.ac.za/ilifu-research-cloud-infrastructure/}}.
To streamline the processing of our \HI data, we used the ThunderKAT \HI Pipeline\footnote{\url{http://github.com/tremou/thunderkat_hi_pipeline.git}}.
Simultaneously, we used \textsc{CARTA} \citep[Cube Analysis and Rendering Tool for Astronomy;][]{Comrie_etal_2024_SW_CARTA} to interrogate the data.

The ThunderKAT \HI Pipeline has three stages, each with its own bash script that employs several \textsc{Python} scripts.

At a high level, the first stage of the pipeline uses \textsc{CASA} \citep[Common Astronomy Software Applications;][]{CASA_Team_etal_2022_SW_CASA} to retrieve the data for specified fields from the full observation measurement set and create separate files for each source.
The pipeline then undoes previously applied flags to ensure \HI spectral lines are not erroneously flagged as radio-frequency interference.
The measurement set for each field is then converted into the \texttt{fits} format required for the \textsc{Miriad} software \citep{Sault_etal_1995_SW_Miriad} used in the next stage.

The second and most computationally intensive stage begins with data pre-processing, and a region is defined to search for the position of the peak continuum emission.
The target and calibrators' fields are defined, the reference antenna is set, and basic flagging is done.
The \HI spectral line frequency is added to the header information to convert frequency to velocity.
Bandpass and gain calibrations are applied to the target field, and spectral cubes are made and cleaned for the target and defined reference source(s).
A second-order polynomial is then fitted to the broadband radio continuum emission of each source and subtracted in frequency space to remove the continuum emission.
The resulting residuals are used to create image cubes for each source, from which the spectra are extracted and written to ASCII files, ready for the final stage.

The third stage plots the target and reference spectra.
The noise in each channel is used to give an estimation of absorption uncertainties in the velocity bins.
In the event that there are multiple target or reference spectra to be combined, these are ``stacked'' to create weighted mean spectra and increase the signal-to-noise ratio (SNR).
The absorption in each velocity bin is calculated from each spectrum weighted according to the inverse square of the noise.

\subsection{Hubble Space Telescope near-UV spectroscopy} \label{sec:data_NUV}

We obtained high resolution near-UV spectroscopy with the Space Telescope Imaging Spectrograph \citep[STIS;][]{Woodgate_etal_1998_HST_STIS_Design} aboard the Hubble Space Telescope (HST) in early October 2023 ($\rm MJD\sim 60219$) during the outburst \citep[program ID 16489;][]{Castro_Segura_etal_2020_HST_Proposal_16489C_Outflow_Legacy_Accretion_Survey}.
We used E230M gratings with 200\,s followed by 220\,s exposures at the central wavelengths 1978\,\angstrom and 2707\,\angstrom respectively to cover the region $\lambda\lambda\simeq1800 \text{--} 3200$\,\angstrom with a resolving power of $R=30,\!000$.
The data were reduced using the HST pipeline {\sc calstis}\footnote{Provided by The Space Telescope Science Institute\\(\url{https://github.com/spacetelescope})}.

\section{Results} \label{sec:results}

\subsection{Radio} \label{sec:results_radio}

\begin{figure*}[t]
    \centering
    \includegraphics[width=0.99\textwidth]{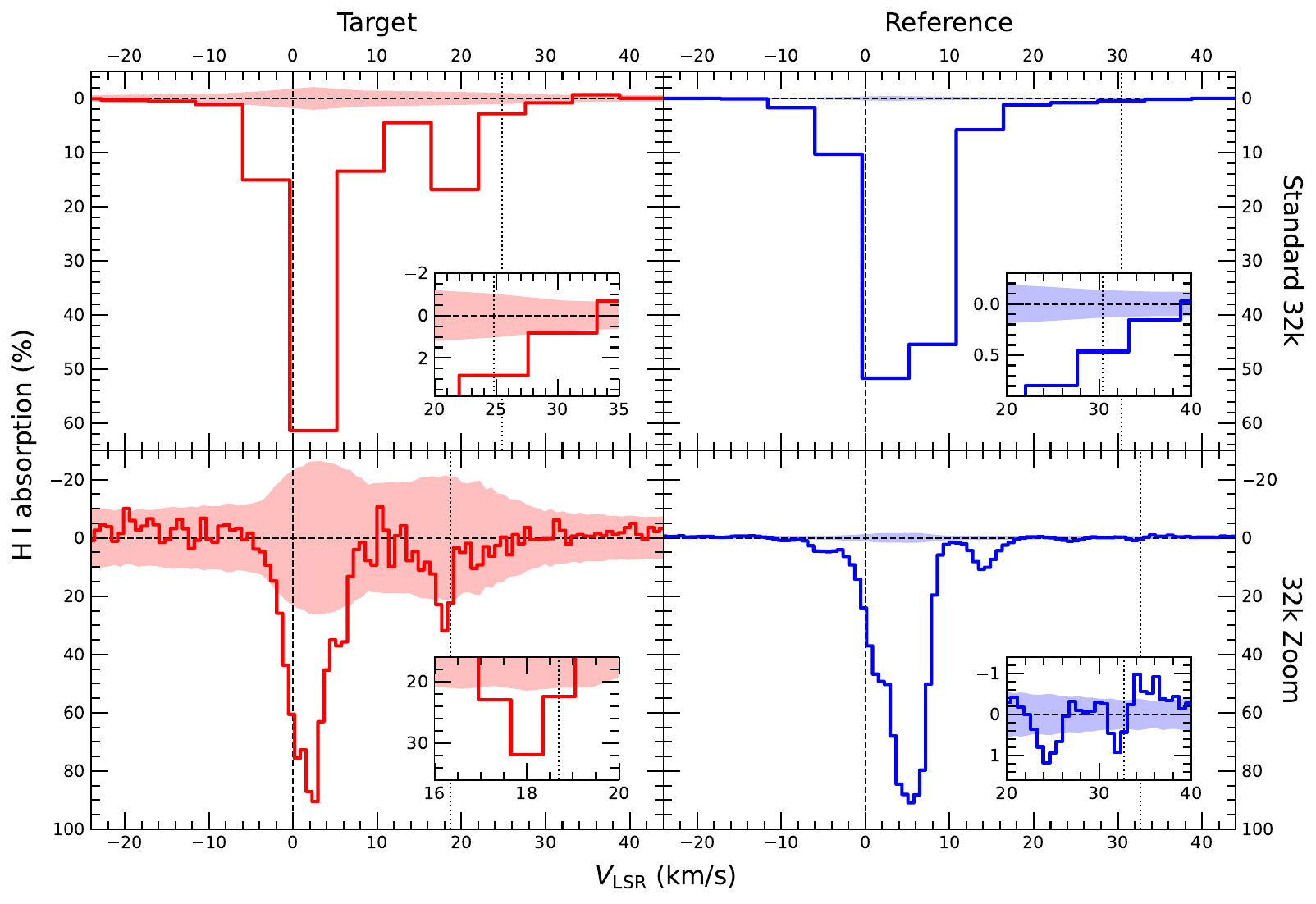}
    \caption{
        A consolidated view of the \HI spectra for both the target, \targ (red, left), and our reference source, \scal (blue, right).
        We observed no significant ($>\!3\sigma$) absorption outside the velocities chosen as the $x$-axis range.
        The $y$-axis represents the \HI absorption percentage.
        The dashed lines indicate the $x$- and $y$-axes origins.
        The vertical dotted lines indicate the maximum velocity at which significant absorption is observed.
        For \targ, these are
            $24.8 \pm 2.8$ and
            $18.7 \pm 0.4$\,\kps.
        For \scal, these are
            $30.4 \pm 2.8$ and
            $32.7 \pm 0.4$\,\kps.
        The top-left plot is the 32k-S spectrum for \targ from 14-10-2023, while the bottom-left plot is the mean weighted \targ spectrum achieved by combining both 32k-Z spectra from Figure \ref{fig:HI_Spectra_SwiftJ1727}.
        We generated the 32k-Z (32k-S) spectra for \scal from observations on 06-09-2023 (14-10-2023) with the peak \scal flux density measured to be $6.1 \pm 0.1$\,Jy ($6.0 \pm 0.1$\,Jy), providing the significantly greater SNR compared to that of the \targ observations.
        }
    \label{fig:HI_Spectra_Comparison}
\end{figure*}

\subsubsection{High resolution 32k zoom mode} \label{sec:results_zoom}

Both \HI spectra from our two 32k-Z observations are displayed in the bottom two plots of Figure \ref{fig:HI_Spectra_SwiftJ1727}. Significant ($>\!3\sigma$) \HI absorption towards \targ is observed out to an estimated maximum LSR velocity $v_{\rm{LSR}} = 18.7 \pm 0.4$\,\kps as shown in the inset for the the mean weighted 32k-Z spectrum at the bottom-left of Figure \ref{fig:HI_Spectra_Comparison}, using half the bin width as the velocity uncertainty.
The maximum absorption is observed to be $\approx 90\%$.

\subsubsection{Standard 32k mode} \label{sec:results_std}

The effect of \targ's variable luminosities over the period of observations is seen in the differing SNR between our \HI spectra from 27-08-2023 and 14-10-2023 shown in Figure \ref{fig:HI_Spectra_SwiftJ1727}.
For 14-10-2023 we measured a \targ peak flux density that was more than eight times greater than either of our 32k-Z observations due to radio flaring, as observed by \citet{Miller-Jones_etal_2023_ATel_16271_Swift_J1727_Radio_quenching_subsequent_flaring}.
The 14-10-2023 spectrum shows \HI absorption at greater velocities than the 32k-Z spectra, albeit with greater bin widths.
We therefore update our estimate of the maximum velocity of significant \HI absorption towards \targ to $v_{\rm{LSR}} = 24.8 \pm 2.8$\,\kps as seen in the top-left inset in Figure \ref{fig:HI_Spectra_Comparison}.
The maximum absorption is observed to be $\approx 61\%$ for the 32k-S spectra, which is smaller than that of the 32k-Z spectra as the absorption is averaged over the wider velocity bin width.

\subsubsection{Reference source} \label{sec:results_reference}

Figure \ref{fig:HI_Spectra_Comparison} compares \HI absorption spectra towards \targ and \scal.
The latter is consistently much brighter, with flux densities exceeding 6 Jy.
It also exhibits \HI absorption to greater velocities (i.e., $32.7 \pm 0.4$\,\kps in the 32k-Z spectrum), which is in line with \scal being extragalactic.
With an extragalactic point of comparison that is nearby in terms of sky location and with \HI absorption to greater velocities, we infer that \targ is closer than the tangent point, and use the near kinematic distance as a lower limit.

\begin{figure}[h!]
    \centering
    \includegraphics[width=0.99\columnwidth]{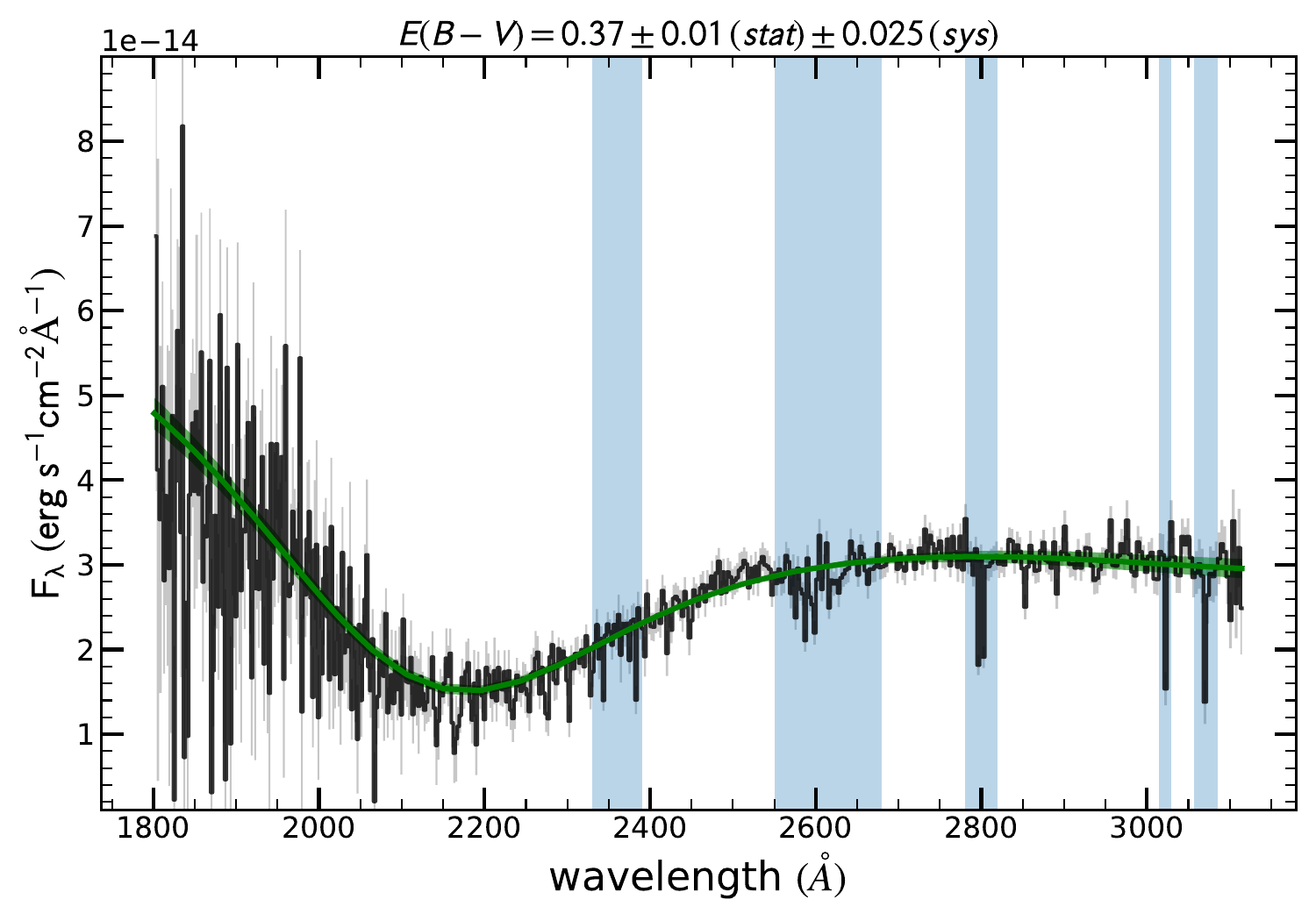}
    \caption{The near-UV spectrum of \targ as seen by HST/STIS (black), with a reddened power-law fit (green) to the data (Castro Segura et al. in prep.) The uncertainty around the fit is indicated as a shaded region. The vertical shaded regions were masked during the fit to avoid Fe and Mg lines that may influence the fit.}
    \label{fig:E(B-V)}
\end{figure}

\subsection{Near-UV} \label{sec:results_NUV}

We determined the line of sight extinction to \targ using the \citet{Cardelli_Clayton_Mathis_1989_Relationship_IR_optical_UV_extinction} reddening law to fit a reddened power-law representing the outer accretion disk to the near-UV spectrum as displayed in Figure \ref{fig:E(B-V)}.
To estimate the errors we performed a Monte Carlo simulation yielding $\EBmV = \EBmVNUV$, corresponding to $A_V \simeq 1.15$ per Equation \ref{eqn:EBmV_to_AV} with $R_V = 3.1$.
Given we use $r$-band magnitudes when calculating Equation \ref{eqn:distance_modulus}, we derive the $r$-band extinction using $A_r = 2.271\,\EBmV = 0.84 \pm 0.06$ \citep[for Pan-STARRS1;][]{Schlafly_Finkbeiner_2011_EBmV_with_SDSS}.

\section{Discussion} \label{sec:discussion}

\subsection{Kinematic distance constraint} \label{sec:discussion_kd}

Absorption features in the final \HI spectra informed our use of the KDCT to generate kinematic distance estimates.
We clearly observe greater velocities towards \scal than \targ, as shown in Figure \ref{fig:HI_Spectra_Comparison}.

We used our maximum \HI absorption velocity and its uncertainty as inputs to the KDCT \citep{Wenger_etal_2018_Kinematic_Distances_MC}, using their Method C and applying the revised solar motion parameters from \citet{Reid_etal_2014_Trig_Parallax_HMSFRs_Structure_Kinematics_MW} and the rotation curve from \citet{Reid_etal_2019_Trig_Parallax_HMSFRs_MW_Our_View}.
The total revised LSR velocity uncertainty includes measurement and systematic uncertainties such as non-circular streaming motions.

We amended the KDCT source code to resample the input and Galactic rotation curve parameters $10^7$ times to minimise Monte Carlo error.
Results include Monte Carlo samples of 
    $d_{\rm{near}}$,
    $d_{\rm{tan}}$ (tangent point kinematic distance), and
    $v_{\rm{LSR,tan}}$ (tangent point LSR velocity).
We estimate and report the median values as best estimates and the boundaries of the highest density 68\% interval as the uncertainties on these quantities.
We repeated this process using inputs from 32k-Z and 32k-S \HI absorption spectra.

The quantities $d_{\rm{tan}}$ and $v_{\rm{LSR,tan}}$ depend only on the source location.
For \targ, we therefore estimate $d_{\rm{tan}} = \dtanstd$\,kpc and $v_{\rm{LSR,tan}} = \vtanstd$\,\kps.
As $d_{\rm{near}}$ depends on the input velocity, we use our higher-SNR result of $24.8 \pm 2.8$\,\kps from our 32k-S spectra to estimate $d_{\rm{near}} = \dnearstd$\,kpc.

\subsubsection{Caveats} \label{sec:discussion_caveats}

\targ has a longitude close to the Galactic Centre (GC) and a relatively high Galactic latitude, leading to larger systematic uncertainties on the kinematic distance.

Regions within $15^{\circ}$ in Galactic longitude from the GC feature increased \HI emission from the GC \citep{Kalberla_Kerp_2009_MW_HI}, which leads to higher sky temperatures around the \HI line and hence increased uncertainty in each spectral channel.
At these longitudes, the motion of objects intersected by the line of sight is mostly perpendicular to the line of sight.
Distances are thus inferred from a smaller spread in circular rotation velocities and subject to larger uncertainties, and so this region was excised from the study of \citet{Wenger_etal_2018_Kinematic_Distances_MC}.

More recently, \citet{Hunter_etal_2024_Testing_Galactic_kinematic_distances} conducted numerical 2D hydrodynamical simulations to account for potential causes of Milky Way deviations from axisymmetry and gas cloud deviations from the circular rotation curve.
The authors categorise these deviations into: (i) random fluctuations around the average streaming motions that do not change the average velocity; and (ii) systematic changes in streaming velocity due to non-axisymmetry, such as spiral arms and the Galactic bar.
The authors then define regions of their simulated Milky Way Galaxy where the discrepancy between the kinematic and true distance is significant ($> \! 27\%$).
Within $2 \! < \! d \! < \! 5$\,kpc, the longitude for \targ appears to correspond to a median absolute relative kinematic distance error of $\vert d_{\rm{k}} - d_{\rm{true}} \vert / d_{\rm{true}} \approx 63\%$.
For $5 \! < \! d \! < \! 10$\,kpc, this error reduces to approximately $12 \pm 8\%$.
We use the greater 63\% error above to expand the $1\sigma$ uncertainty on our measured kinematic distance, which becomes $\dnearunc$\,kpc.

We observe that minor changes in Galactic longitudes close to the GC have major impacts on the values of $d_{\rm{near}}$ calculated using the KDCT.
Specifically, \scal has a maximum \HI absorption velocity that is more than 30\% greater than that of \targ, however its larger Galactic longitude of $l \approx 12^{\circ}$ results in a similar predicted value of $d_{\rm{near}}$.

Our conservative lower limit of $\dnearunc$\,kpc by accounting for longitudinal effects may therefore be reasonable, given the high Galactic latitude of \targ and thus diminishing \HI density along the lines of sight towards \targ and \scal.

\subsubsection{Scale height of Galactic H\,{\footnotesize I}} \label{sec:discussion_height}

The kinematic distance method works best for sources located in the GP where $b \! \approx \! 0$.
\citet{Wenger_etal_2018_Kinematic_Distances_MC} assumed a latitude of $b \! = \! 0$ and only use latitude to correct the LSR velocity with updated solar motion parameters.
Having used 2D simulations, \citet{Hunter_etal_2024_Testing_Galactic_kinematic_distances} assumed that the gas is integrated along the $z$-axis (vertically), and that the acceleration of the gas due to the Galactic potential is computed as if the gas lies in the GP with Galactic elevation, $z$, equal to zero.
High Galactic latitudes correspond to greater Galactic elevations where less gas and other matter reside.
Observed absorption features are primarily due to the gas clouds that are closer to us, as evidenced by the absence of detectable \HI absorption out to the tangent point towards \scal.
The impact of the Galactic latitude for \targ will therefore lead to, if anything, an underestimation of the distance.

We do not observe \HI absorption to the tangent point velocity of $\vtanstd$\,\kps in any of our spectra; only up to a maximum velocity $32.7 \pm 0.4$\,\kps in the direction of \scal.
This suggests that the line of sight has not intersected \HI clouds at greater distances due to $z$ increasing and \HI density decreasing with distance.
The distance lower bound of $d_{\rm{near}} = \dnearunc$\,kpc corresponds to a GP elevation of $z \approx 650 \pm 410$\,pc.

\citet{Kalberla_Kerp_2009_MW_HI} suggest that the scale height of Milky Way's \HI disk is approximately 150 pc at $R=0$.
The flaring and warping of the \HI disk discussed by the authors would not be significant at the location of \targ given it resides within the solar circle.
More recently, \citet{Rybarczyk_Wenger_Stanimirovic_2024_Vert_Dist_HI_Clouds_21cm_Abs_High_Gal_Lat} showed the Gaussian-distributed thickness of the cold neutral medium in the solar neighbourhood, $\sigma_z$, to be no more than $\sim 150$\,pc.
It can therefore be expected that the majority of \HI clouds along the lines of sight towards our sources will be contained within a few multiples of this $\sigma_z$.
\citet{Rybarczyk_Wenger_Stanimirovic_2024_Vert_Dist_HI_Clouds_21cm_Abs_High_Gal_Lat} also found that \HI features at $|b|>5^{\circ}$ trace primarily local structures, within 2\,kpc.
From their Figure 7, we see that the maximum \HI absorption that we observed toward \targ and \scal falls into a region of outliers in Galactic position-velocity space, lending credence to our decision to impose additional uncertainty on our kinematic distance lower limit.

\subsubsection{Feasibility of H\,{\footnotesize I} absorption XRB distances} \label{sec:discussion_HIfeasibility}

Given the constraints made possible from a small number of XRB observations with MeerKAT, kinematic distances via \HI absorption studies have the potential to form the basis of a rapid, routine, and reasonably accurate method for placing informative limits on the distances to Galactic transients, especially those situated within the GP and/or further from the GC than \targ.
This method will become increasingly powerful when used in conjunction with Square Kilometre Array observations of sufficiently bright XRBs in outburst, albeit with the caveats discussed above.

\subsection{$\EBmV$ distance constraint} \label{sec:discussion_EBmVdist}

\subsubsection{$\EBmV$ validation} \label{sec:discussion_EBmVvalidation}

\begin{figure*}[t]
    \centering
    \includegraphics[width=0.90\textwidth]{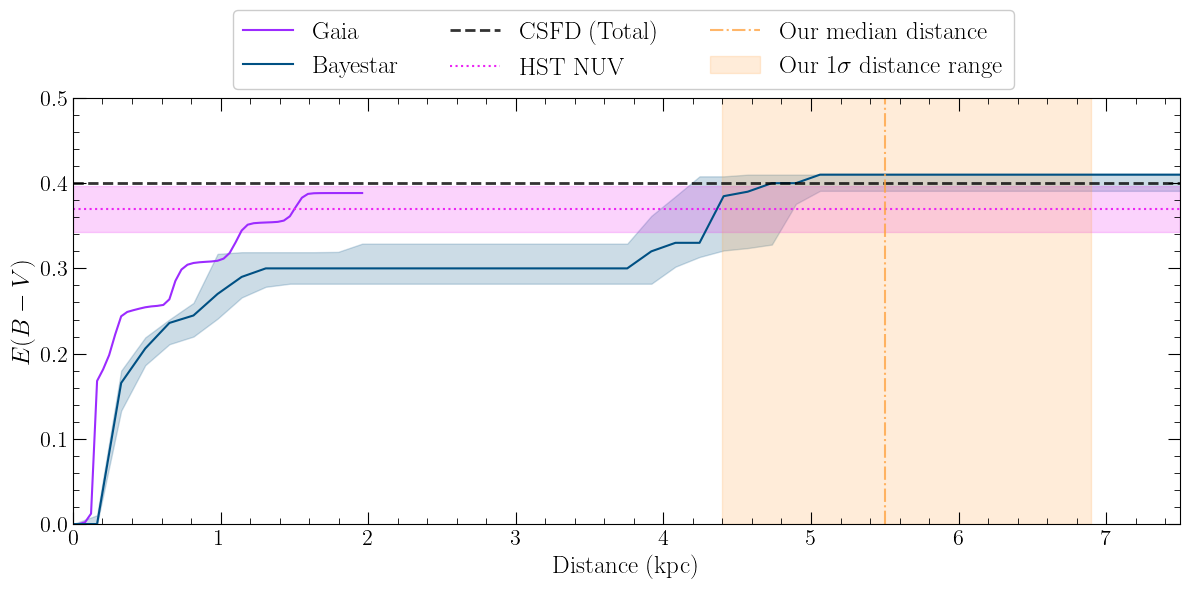}
    \caption{
        Estimates of $\EBmV$ for distances between $0 \text{--} 7.5$\,kpc along the line of sight to \targ.
        The 3D Galactic dust maps include {\em Bayestar} \citep[blue line;][]{Green_etal_2019_3D_dust_map_Gaia_Pan-STARRS1_2MASS} and {\em Gaia} \citep[purple line;][]{Edenhofer_etal_2024_pc_scale_3D_dust_map}. We also use an update of the 2D ``SFD'' dust map \citep{Schlegel_Finkbeiner_David_1998_SFD_dust_map_via_IR_emission_for_reddening_and_CMBRFs} to remove extragalactic contamination that is referred to as the ``corrected SFD'' \citep[CSFD; grey dashed line;][]{Chiang_2023_Corrected_SFD_CSFD_dust_map_minimal_extragalactic_contamination}.
        To investigate the likely maximum $\EBmV$ values within this distance range, we use CSFD; as this only reports the total value for $\EBmV$ along the line of sight, it serves as a good estimate of the maximum $\EBmV$ in the direction of \targ at high distances.
        From this we derive a maximum $\EBmV \approx 0.4$.
        Despite each having significant caveats and the disagreement between {\em Bayestar} and {\em Gaia} in how gas is distributed along the line of sight, both appear to accumulate to roughly this maximum value.
        Lastly, our estimate of $\EBmV$ (pink dotted line) is derived from our near-UV observations.
        We use this estimate to conclude that values of $\EBmV \gg \EBmVuplim$ may not be reliable, and suggest a distance to \targ of $\dfinal$\,kpc with $1\sigma$ uncertainties, which are included for reference (orange dot-dash line and shaded region respectively).
    }
    \label{fig:extinction}
\end{figure*}

To first validate our result of $\EBmV = \EBmVNUV$ as derived from near-UV observations in Section \ref{sec:results_NUV}, we compute the change in $\EBmV$ along the line of sight to \targ using Galactic dust maps.
We present these results in Figure \ref{fig:extinction}.
Our results imply that $\EBmV$ cannot be much greater than 0.4, as we run out of dust along the line of sight (but note the spatial resolution of dust maps could be a limiting factor).
In fact, our best value for $\EBmV$ is lower than all those estimated by \mata.

We note that while there is no evidence for intrinsic variability in $\NH$, the values used in studies of \targ have varied from as low as $1.2 \pm 0.2 \times 10^{21}\,{\rm cm^{-2}}$ \citep{Chatterjee_etal_2024_Swift_J1727_Insight-HXMT_view_outburst} to as high as $(4.1 \pm 0.1) \times 10^{21}\,{\rm cm^{-2}}$ \citep{Draghis_etal_2023_ATel_16219_Swift_J1727_prelim_specfit_QPO_evo_NICER_obs} with other values in between \citep{O'Connor_etal_2023_ATel_16207_Swift_J1727_NICER_detection, Peng_etal_2024_Swift_J1727_NICER_NuSTAR_Insight-HXMT_views, Svoboda_etal_2024_Swift_J1727_X-ray_pol_drop_in_soft_state}.
These discrepancies are likely due to the instrumental systematics and/or modelling choices differing from study to study.

\mata used $(3.2 \pm 0.9) \times 10^{21}\,{\rm cm^{-2}}$, derived as the mean $\NH$ value from \citet{O'Connor_etal_2023_ATel_16207_Swift_J1727_NICER_detection} and \citet{Draghis_etal_2023_ATel_16219_Swift_J1727_prelim_specfit_QPO_evo_NICER_obs}.
The latter was estimated using the \textsc{tbabs} X-ray absorption model and \citet{Wilms_etal_2000_ISM_X-ray_Absorption} abundances of elements.
Both were derived by fitting observations made in late August 2023, when the source was in the rising hard/hard-intermediate state.

We instead use $(2.4 \pm 0.1) \times 10^{21}\,{\rm cm^{-2}}$ per \citet{Svoboda_etal_2024_Swift_J1727_X-ray_pol_drop_in_soft_state}, which was derived using NICER, NuSTAR, and IXPE observations made during the soft state.
The broad spectral coverage of this data set and the well-characterised disk blackbody spectrum during the soft state allows for a more robust estimate of $\NH$ than those derived from a more restricted bandpass or made during the hard or intermediate states.
This value is in agreement with the HI4PI Survey's measured $\NH$ in the direction of \targ \citep{HI4PI_Collaboration_etal_2016_Fullsky_H_I_survey_EBHIS_GASS}, and reasonable agreement with the $\NH$ values presented by \citet{Chatterjee_etal_2024_Swift_J1727_Insight-HXMT_view_outburst} and \citet{Peng_etal_2024_Swift_J1727_NICER_NuSTAR_Insight-HXMT_views}.

We also use an alternative relation, $\NH ({\rm cm}^{-2}) = (2.87 \pm 0.04) \times 10^{21} A_V {\rm\,(mag)}$ \citep[][Equation 11]{Zhu_etal_2017_Galactic_gas_to_extinction_and_distribution}, rather than that of \citet{Guver_etal_2009_Optical_extinction_hydrogen_column_density}.
This more recent relation is a more appropriate choice for estimating extinction, given it was determined by fitting the whole sample of \citet{Wilms_etal_2000_ISM_X-ray_Absorption} abundances, which include XRBs.
However, we note there is likely additional uncertainty due to the scatter observed in this relation \citep[][Figure 9(a)]{Zhu_etal_2017_Galactic_gas_to_extinction_and_distribution}.
Assuming $R_V=3.1$, we use Monte Carlo techniques and Equation \ref{eqn:EBmV_to_AV} to instead estimate $\EBmV = \EBmVNH$ along the line of sight to \targ.
This result, while dependent upon the choice of $\NH$ and with likely greater uncertainty, are in agreement with our findings based on near-UV data and Galactic dust maps that $\EBmV \lesssim 0.4$.

\subsubsection{Suggested distance to \targf}

\begin{figure}[h!]
    \centering
    \includegraphics[width=0.99\columnwidth]{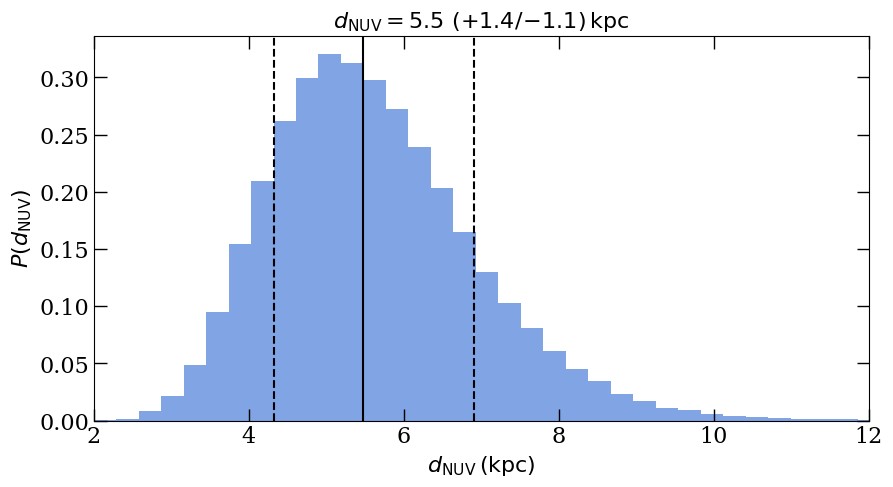}
    \caption{The distance posterior distribution produced from Monte Carlo calculations using the near-UV constraint on $\EBmV$.
    The solid vertical line represents the 0.5000 quantile and the dashed vertical lines represent the 0.1585 and 0.8415 quantiles.}
    \label{fig:distance}
\end{figure}

With $A_r = 0.84 \pm 0.06$ per Section \ref{sec:results_NUV}, the remaining parameter values required for Equation \ref{eqn:distance_modulus} are the absolute and apparent $r$-band magnitudes, $M_r$ and $m_r$ respectively.
\matat determine the donor star type to be K4($\pm$1)V with $M_r = 6.6 \pm 0.5$\,mag and $m_r = 21.13 \pm 0.05$\,mag.
Using these values produces the distance posterior distribution that we present in Figure \ref{fig:distance} and a distance estimate of $\dfinal$\,kpc, under the assumption that the donor star is a regular K4($\pm$1)V star that has not lost significant mass during the binary evolution.
In such a case, the distance would be reduced, implying reduced values of $\EBmV$ and $\NH$.

\subsubsection{Nature of the donor star} \label{sec:discussion_donor}

Using our distance estimate and the results from \citet{Wood_etal_2025_Swift_J1727_Ejection_Transient_Jets_Time_Dependent_Visibility_Modelling}, we define an upper limit on the inclination angle, $i$, for \targ of $i \lesssim 69^{\circ}$.
To define a conservative lower limit on $i$, we assume an upper limit of $20\,{\rm M_{\odot}}$ for the black hole mass.
It is likely that the black hole is actually much smaller, as discussed with reference to high natal kick velocity in Section \ref{sec:discussion_implications}.
We also use a maximum-entropy skew-normal distribution for the K4($\pm$1)V donor star mass such that $1\sigma$ is contained within the range $0.70 \text{--} 0.78\,{\rm M_{\odot}}$ and $\mu \approx 0.73\,{\rm M_{\odot}}$.
These mass values allow us to use the mass function provided by \matat,
\begin{equation}
    \label{eqn:mass_function}
    f(M) \equiv \frac{M_1 \sin^3 i}{(1 + M_2/M_1)^2} = 2.77 \pm 0.09 {\rm M_{\odot}},
\end{equation}
to estimate a $1\sigma$ lower limit of $i \geq 32.0\,\pm\,0.4^{\circ}$.

\matat obtained an upper limit of $v \sin i < 52$\,\kps for the rotational broadening of the donor star, but recommend a more conservative $3\sigma$ upper limit of $102$\,\kps as this constraint is not especially strong.
Values of $v \sin i > 102$\,\kps are therefore unlikely, and would require $M_2 > 1 M_{\odot}$.

Given the binary system's orbital period, $\Porb$, and assuming that the donor star is tidally locked, the donor star's radius can be calculated as
\begin{equation}
    \label{eqn:donor_radius}
    R_2 = \frac{\Porb(v \sin i)}{2 \pi \sin i}.
\end{equation}
\matat directly measured $\Porb = 10.8038 \pm 0.0010$\,hrs.
Combining this with the $v \sin i = 27^{+25}_{-21}$\,\kps posterior distribution from \matat and a uniform distribution of $\cos i \sim U(\cos 69^{\circ},\cos 32^{\circ})$ results in an $R_2$ distribution where $R_2 \ll 1.0\,{\rm R_{\odot}}$, whereas K4($\pm$1)V stars typically have radii of $R \approx 1\,{\rm R_{\odot}}$.

\citet{Paczynski_1967_Evolution_Close_Binaries_Wolf-Rayet_Stars} relates the XRB orbital period and the mass and radius of a donor star undergoing Roche lobe overflow as
\begin{equation} \label{eqn:Paczynski}
    R_L \approx (2 G M_2)^{1/3} (\Porb / 9 \pi)^{2/3}
\end{equation}
where $R_L$ is the Roche lobe radius, and $M_2$ is the donor star (secondary) mass.
Taking $R_L = R_2$ as derived using the $v \sin i$ posterior distribution from \matat results in an $M_2$ distribution that also favours extremely small, unphysical values ($M_2 \ll 0.1\,{\rm M_{\odot}}$).

Donor stars of BH low-mass XRBs may be significantly evolved \citep{Podsiadlowski+Rappaport+Han_2003_Formation_Evolution_BH_Binaries, Fragos+McClintock_2015_Origin_BH_Spin_LMXBs} and in Roche lobe overflow.
However, there is no current evidence from the donor star spectral analysis by \matat that it is significantly evolved or stripped, which would impact the distance estimate.
Only with future, higher resolution spectroscopic observations could we gain further information into the nature of the \targ donor star.

We therefore infer that $v \sin i$ is unlikely to be as low as the \matat posterior distribution might suggest.
Assuming co-rotation and Roche lobe overflow, we again combine equations \ref{eqn:donor_radius} and \ref{eqn:Paczynski}.
Then, using $\Porb$ from \matat and our aforementioned $M_2$ skew-normal and $\cos i$ uniform distributions, we calculate $v \sin i$ to be in the $3\sigma$ range $60 \text{--} 112$\,\kps.
This is roughly consistent with \matat $3\sigma$ upper limit $v \sin i \lesssim 102$\,\kps, given it is dependent upon conservative limits on $i$, and further contradicts the abundance of low values in the \matat $v \sin i$ posterior.

\subsection{Implications} \label{sec:discussion_implications}

\subsubsection{Natal kick velocity and birth mechanism} \label{sec:discussion_natalkick}

Using their estimated distance of $3.4 \pm 0.3$\,kpc, along with the proper motion of \targ, \matat derived a median and $1\sigma$ confidence level in the potential natal kick velocity of $v_{\rm{kick}} = 210^{+40}_{-50}$\,\kps.
We use our suggested distance of $\dfinal$\,kpc to provide an updated natal kick velocity of $\kickv$\,\kps \citep{Atri_etal_2019_BH_XRB_natal_kick_velocity_distribution}\footnote{\url{https://github.com/pikkyatri/BHnatalkicks}}.
Our revised distance also corresponds to Galactic elevations of $z \approx 0.8 \text{--} 1.2$\,kpc.
We can compare this to XTE J1118+480, for which \citet{Gualandris_etal_2005_XTE_J1118+480_Asymmetric_Natal_Kick} presented an asymmetric natal kick with a similar velocity of $183 \pm 31$\,\kps and a Galactic elevation of $1.9 \pm 0.4$\,kpc.
\citet{Atri_etal_2019_BH_XRB_natal_kick_velocity_distribution} estimated a similar natal kick velocity for XTE J1118+480 and particularly high velocity kicks for other systems, such as 4U 1543$-$475, GS 1354$-$64, and SAX J1819.3$-$2525, and suggested that these are indicative of asymmetrical supernovae explosions as the likely birth mechanism.
Applying this to \targ, it is possible that it could have been born within the Galactic disk, and propelled by a natal kick to its present-day Galactic elevation. It is also possible that the large natal kick has caused a misalignment between the BH spin and the accretion disk for \targ  \citep{Maccarone_2002_Microquasar_jet_misalignment,Martin_etal_2008_GRO_J1655-40_Alignment_timescale_microquasar}. However, no evidence has been observed to date of precession of the jet axis \citep{Wood_etal_2024_Swift_J1727_Largest_Resolved_Jet_XRB,Wood_etal_2025_Swift_J1727_Ejection_Transient_Jets_Time_Dependent_Visibility_Modelling}.

\subsubsection{Eddington luminosity fraction} \label{sec:discussion_ELF}

\citet{Zdziarski+Wood+Carotenuto_2025_Swift_J1727_Modeling_Extended_Emission_Compact_Jets} calculate an unabsorbed bolometric flux of $F_{\rm bol} \approx 5 \times 10^{-7}\,{\rm\,erg\,cm^{-2}\,s^{-1}}$ for \targ, as informed by the peak hard state flux from \citet{Liu_etal_2024_Swift_J1727_Broadband_X-ray_Spectral_Properties_Rising_Outburst}.
At a distance of $\dfinal$\,kpc, this would correspond to a luminosity of $L_{\rm bol} = 1.8^{+1.1}_{-0.7} \times 10^{39}\,\rm{erg\,s^{-1}}$.
Assuming a nominal BH mass of $10\,M_{\odot}$, $L_{\rm Edd} \approx 1.3 \times 10^{39}\,\rm{erg\,s^{-1}}$, meaning an ELF of $L_{\rm bol} / L_{\rm Edd} \approx 1.4^{+0.9}_{-0.6}$ in this case.
Black holes with higher predicted natal kicks are expected to be less massive \citep[e.g.;][]{Belczynski+Bulik_2002_GRS_1915_Formation_Evolution, Maccarone_etal_2020_4U_1957+11_Upper_limit_jet_power}, yet reducing the BH mass would only increase this already-large ELF.
Therefore, any distance within our $1\sigma$ uncertainties of \drange is set to contravene expectations from \citet{Zdziarski+Wood+Carotenuto_2025_Swift_J1727_Modeling_Extended_Emission_Compact_Jets} who note that the highest observed luminosities of low-mass XRBs in the hard or hard-intermediate states reside in the regime $L\,\lesssim\,0.3\,L_{\rm Edd}$.
\citet{Zdziarski+Wood+Carotenuto_2025_Swift_J1727_Modeling_Extended_Emission_Compact_Jets} also noted that \targ underwent a soft-to-hard state transition at very low flux in the period February--April 2024, at $F_{\rm bol} \approx 10^{-8}\,{\rm\,erg\,cm^{-2}\,s^{-1}}$.
With this flux, our same distance corresponds to a soft-to-hard state transition ELF of $0.029^{+0.016}_{-0.010}$ assuming a BH mass of $10\,M_{\odot}$, which is more in line with expectations of $1 \text{--} 4\%$ \citep{Kalemci_etal_2013_BHT_distance_state_transitions,VahdatMotlagh+Kalemci+Maccarone_2019_State_transition_luminosities_Galactic_BHT_outburst_decay}.
These ELFs imply that the distance to \targ is likely on the lower end of our $1\sigma$ range.
However, no single distance can satisfy the expectations on both the peak hard state luminosity and soft-to-hard state transition luminosity.
While ELFs from soft-to-hard state transitions appear to be steadier \citep{Kalemci_etal_2013_BHT_distance_state_transitions}, we caution that both the above ELFs are merely indicative and are unlikely to be reliable without a more accurate constraint on the BH mass.

\section{Conclusions} \label{sec:conclusions}

Using \HI absorption data from MeerKAT observations of the outburst of \targf, we determine a maximum absorption velocity of $24.8 \pm 2.8$\,\kps.
The higher-velocity absorption seen towards the extragalactic reference source \scalf allows us to use the near kinematic distance as a lower bound of $\dnearunc$\,kpc, accounting for the systematic uncertainty due to the low Galactic longitude of the source.
However, its high Galactic latitude likely implies that we run out of \HI clouds along the line of sight.

Making use of a near-UV spectrum of \targf as observed by the Space Telescope Imaging Spectrograph aboard the Hubble Space Telescope, we measure the colour excess or reddening to be $\EBmV = \EBmVNUV$.
This is significantly lower than previous constraints on $\EBmV$ from \citet{MataSanchez_etal_2024_Swift_J1727_Inflows_Outflows,MataSanchez_etal_2025_Swift_J1727_Dynamically_Confirmed_BH}, but it is in good agreement the maximum value derived from Galactic dust maps along this line of sight.

We use $\EBmV$ to determine the $r$-band extinction, $A_r$, and combine this with donor star $r$-band magnitudes provided by \citet{MataSanchez_etal_2025_Swift_J1727_Dynamically_Confirmed_BH} for a K4($\pm$1)V main sequence donor star.
Under this assumption, we subsequently present $\dfinal$\,kpc as the resulting and likely distance to \targf, which implies a natal kick velocity of $v_{\rm{kick}} = \kickv$\,\kps and suggests a likely formation in a natal supernova.

If the donor star has instead lost significant mass during the binary evolution, this distance would be smaller.
However, the exact evolutionary stage and possible mass loss through accretion of the donor star is unknown.
Further observations to better constrain binary inclination angle and the donor star rotational broadening would allow more accurate determination of the primary and secondary masses and consequently the distance.

\section*{Acknowledgements}

The authors would like to thank James Allison for his development of the ThunderKAT \HI Pipeline.

The authors would also like to thank Ilya Mandel, Ryosuke Hirai, Natasha Ivanova, and Thomas Maccarone for useful discussions.

The MeerKAT telescope is operated by the South African Radio Astronomy Observatory, which is a facility of the National Research Foundation, an agency of the Department of Science, Technology and Innovation.

We acknowledge the use of the ilifu cloud computing facility -- \url{www.ilifu.ac.za}, a partnership between the University of Cape Town, the University of the Western Cape, Stellenbosch University, Sol Plaatje University and the Cape Peninsula University of Technology. The ilifu facility is supported by contributions from the Inter-University Institute for Data Intensive Astronomy (IDIA -- a partnership between the University of Cape Town, the University of Pretoria and the University of the Western Cape), the Computational Biology division at UCT and the Data Intensive Research Initiative of South Africa (DIRISA).

This work made use of the CARTA (Cube Analysis and Rendering Tool for Astronomy) software (\citealt{Comrie_etal_2024_SW_CARTA} -- \url{https://cartavis.github.io}).

This research is based on observations made with the NASA/ESA Hubble Space Telescope obtained from the Space Telescope Science Institute, which is operated by the Association of Universities for Research in Astronomy, Inc., under NASA contract NAS 5--26555. These observations are associated with program 16489. 

Noel Castro Segura acknowledges support from the Science and Technology Facilities Council (STFC) grant ST/X001121/1.

Andrzej Zdziarski acknowledges support from the Polish National Science Center grants 2019/35/B/ST9/03944 and 2023/48/Q/ST9/00138.

Daniel Mata S\'anchez acknowledges support via a Ram\'on y Cajal Fellowship RYC2023-044941.

\facilities{MeerKAT, HST/STIS}.\\

\software{
    \textsc{ArviZ} \citep{Kumar_etal_2019_SW_ArviZ}, 
    \textsc{Astropy} \citep{Astropy_Collaboration_etal_2022_SW_Astropy}, 
    \textsc{CASA} \citep{CASA_Team_etal_2022_SW_CASA}, 
    \textsc{CARTA} \citep{Comrie_etal_2024_SW_CARTA}, 
    \textsc{dustmaps} \citep{Green_2018_SW_dustmaps}, 
    \textsc{Jupyter} \citep{Kluyver_etal_2016_SW_Jupyter}, 
    \textsc{kd} \citep{Wenger_2018_SW_KDCT_Zenodo_DOI_1166001}, 
    \textsc{Matplotlib} \citep{Hunter_2007_SW_Matplotlib}, 
    \textsc{Miriad} \citep{Sault_etal_1995_SW_Miriad}, 
    \textsc{NumPy} \citep{Harris_etal_2020_SW_NumPy}, 
    \textsc{PreliZ} \citep{Icazatti_etal_2023_SW_PreliZ}, 
    \textsc{PyMC} \citep{AbrilPla_etal_2023_SW_PyMC}, 
    \textsc{SciPy} \citep{Pauli_etal_2020_SW_SciPy}, 
    \textsc{stistools} \citep{Hack_etal_2018_SW_stistools}.
    }

\bibliography{References.bib}{}

\begin{thebibliography}{}
\expandafter\ifx\csname natexlab\endcsname\relax\def\natexlab#1{#1}\fi
\providecommand{\url}[1]{\href{#1}{#1}}
\providecommand{\dodoi}[1]{doi:~\href{http://doi.org/#1}{\nolinkurl{#1}}}
\providecommand{\doeprint}[1]{\href{http://ascl.net/#1}{\nolinkurl{http://ascl.net/#1}}}
\providecommand{\doarXiv}[1]{\href{https://arxiv.org/abs/#1}{\nolinkurl{https://arxiv.org/abs/#1}}}

\bibitem[{{Abdulghani} {et~al.}(2024){Abdulghani}, {Lohfink}, \&
  {Chauhan}}]{Abdulghani_etal_2024_Swift_J1727_X-ray_Transition_Luminosities_ELF}
{Abdulghani}, Y., {Lohfink}, A.~M., \& {Chauhan}, J. 2024, \mnras, 530, 424,
  \dodoi{10.1093/mnras/stae767}

\bibitem[{Abril-Pla {et~al.}(2023)Abril-Pla, Andreani, Carroll, Dong,
  Fonnesbeck, Kochurov, Kumar, Lao, Luhmann, Martin, Osthege, Vieira, Wiecki,
  \& Zinkov}]{AbrilPla_etal_2023_SW_PyMC}
Abril-Pla, O., Andreani, V., Carroll, C., {et~al.} 2023, PeerJ Computer
  Science, 9, e1516, \dodoi{10.7717/peerj-cs.1516}

\bibitem[{{Arnason} {et~al.}(2021){Arnason}, {Papei}, {Barmby}, {Bahramian}, \&
  {Gorski}}]{Arnason_etal_2021_Distances_Galactic_XRBs_Gaia_DR2}
{Arnason}, R.~M., {Papei}, H., {Barmby}, P., {Bahramian}, A., \& {Gorski},
  M.~D. 2021, \mnras, 502, 5455, \dodoi{10.1093/mnras/stab345}

\bibitem[{{Astropy Collaboration} {et~al.}(2022){Astropy Collaboration},
  {Price-Whelan}, {Lim}, {Earl}, {Starkman}, {Bradley}, {Shupe}, {Patil},
  {Corrales}, {Brasseur}, {N{\"o}the}, {Donath}, {Tollerud}, {Morris},
  {Ginsburg}, {Vaher}, {Weaver}, {Tocknell}, {Jamieson}, {van Kerkwijk},
  {Robitaille}, {Merry}, {Bachetti}, {G{\"u}nther}, {Aldcroft},
  {Alvarado-Montes}, {Archibald}, {B{\'o}di}, {Bapat}, {Barentsen},
  {Baz{\'a}n}, {Biswas}, {Boquien}, {Burke}, {Cara}, {Cara}, {Conroy},
  {Conseil}, {Craig}, {Cross}, {Cruz}, {D'Eugenio}, {Dencheva}, {Devillepoix},
  {Dietrich}, {Eigenbrot}, {Erben}, {Ferreira}, {Foreman-Mackey}, {Fox},
  {Freij}, {Garg}, {Geda}, {Glattly}, {Gondhalekar}, {Gordon}, {Grant},
  {Greenfield}, {Groener}, {Guest}, {Gurovich}, {Handberg}, {Hart},
  {Hatfield-Dodds}, {Homeier}, {Hosseinzadeh}, {Jenness}, {Jones}, {Joseph},
  {Kalmbach}, {Karamehmetoglu}, {Ka{\l}uszy{\'n}ski}, {Kelley}, {Kern},
  {Kerzendorf}, {Koch}, {Kulumani}, {Lee}, {Ly}, {Ma}, {MacBride}, {Maljaars},
  {Muna}, {Murphy}, {Norman}, {O'Steen}, {Oman}, {Pacifici}, {Pascual},
  {Pascual-Granado}, {Patil}, {Perren}, {Pickering}, {Rastogi}, {Roulston},
  {Ryan}, {Rykoff}, {Sabater}, {Sakurikar}, {Salgado}, {Sanghi}, {Saunders},
  {Savchenko}, {Schwardt}, {Seifert-Eckert}, {Shih}, {Jain}, {Shukla}, {Sick},
  {Simpson}, {Singanamalla}, {Singer}, {Singhal}, {Sinha}, {Sip{\H{o}}cz},
  {Spitler}, {Stansby}, {Streicher}, {{\v{S}}umak}, {Swinbank}, {Taranu},
  {Tewary}, {Tremblay}, {de Val-Borro}, {Van Kooten}, {Vasovi{\'c}}, {Verma},
  {de Miranda Cardoso}, {Williams}, {Wilson}, {Winkel}, {Wood-Vasey}, {Xue},
  {Yoachim}, {Zhang}, {Zonca}, \& {Astropy Project
  Contributors}}]{Astropy_Collaboration_etal_2022_SW_Astropy}
{Astropy Collaboration}, {Price-Whelan}, A.~M., {Lim}, P.~L., {et~al.} 2022,
  \apj, 935, 167, \dodoi{10.3847/1538-4357/ac7c74}

\bibitem[{{Atri} {et~al.}(2019){Atri}, {Miller-Jones}, {Bahramian}, {Plotkin},
  {Jonker}, {Nelemans}, {Maccarone}, {Sivakoff}, {Deller}, {Chaty}, {Torres},
  {Horiuchi}, {McCallum}, {Natusch}, {Phillips}, {Stevens}, \&
  {Weston}}]{Atri_etal_2019_BH_XRB_natal_kick_velocity_distribution}
{Atri}, P., {Miller-Jones}, J.~C.~A., {Bahramian}, A., {et~al.} 2019, \mnras,
  489, 3116, \dodoi{10.1093/mnras/stz2335}

\bibitem[{{Atri} {et~al.}(2020){Atri}, {Miller-Jones}, {Bahramian}, {Plotkin},
  {Deller}, {Jonker}, {Maccarone}, {Sivakoff}, {Soria}, {Altamirano},
  {Belloni}, {Fender}, {Koerding}, {Maitra}, {Markoff}, {Migliari}, {Russell},
  {Russell}, {Sarazin}, {Tetarenko}, \&
  {Tudose}}]{Atri_etal_2020_MAXI_J1820_parallax}
---. 2020, \mnras, 493, L81, \dodoi{10.1093/mnrasl/slaa010}

\bibitem[{{Beardmore} {et~al.}(2016){Beardmore}, {Willingale}, {Kuulkers},
  {Altamirano}, {Motta}, {Osborne}, {Page}, \&
  {Sivakoff}}]{Beardmore_etal_2016_LotR_RotK_X-ray_dust_scattering_rings_V404_Cyg}
{Beardmore}, A.~P., {Willingale}, R., {Kuulkers}, E., {et~al.} 2016, \mnras,
  462, 1847, \dodoi{10.1093/mnras/stw1753}

\bibitem[{{Belczynski} \&
  {Bulik}(2002)}]{Belczynski+Bulik_2002_GRS_1915_Formation_Evolution}
{Belczynski}, K., \& {Bulik}, T. 2002, \apjl, 574, L147, \dodoi{10.1086/342480}

\bibitem[{{Bright} {et~al.}(2023){Bright}, {Farah}, {Fender}, {Siemion},
  {Pollak}, \&
  {DeBoer}}]{Bright_etal_2023_ATel_16228_Swift_J1727_ATA_detection}
{Bright}, J., {Farah}, W., {Fender}, R., {et~al.} 2023, The Astronomer's
  Telegram, 16228, 1.
\newblock \url{https://www.astronomerstelegram.org/?read=16228}

\bibitem[{{Camilo}(2018)}]{Camilo_2018_Afican_star_joins_radio_astronomy_firmament_MeerKAT}
{Camilo}, F. 2018, Nature Astronomy, 2, 594, \dodoi{10.1038/s41550-018-0516-y}

\bibitem[{{Cardelli} {et~al.}(1989){Cardelli}, {Clayton}, \&
  {Mathis}}]{Cardelli_Clayton_Mathis_1989_Relationship_IR_optical_UV_extinction}
{Cardelli}, J.~A., {Clayton}, G.~C., \& {Mathis}, J.~S. 1989, \apj, 345, 245,
  \dodoi{10.1086/167900}

\bibitem[{{CASA Team} {et~al.}(2022){CASA Team}, {Bean}, {Bhatnagar}, {Castro},
  {Donovan Meyer}, {Emonts}, {Garcia}, {Garwood}, {Golap}, {Gonzalez Villalba},
  {Harris}, {Hayashi}, {Hoskins}, {Hsieh}, {Jagannathan}, {Kawasaki},
  {Keimpema}, {Kettenis}, {Lopez}, {Marvil}, {Masters}, {McNichols},
  {Mehringer}, {Miel}, {Moellenbrock}, {Montesino}, {Nakazato}, {Ott}, {Petry},
  {Pokorny}, {Raba}, {Rau}, {Schiebel}, {Schweighart}, {Sekhar}, {Shimada},
  {Small}, {Steeb}, {Sugimoto}, {Suoranta}, {Tsutsumi}, {van Bemmel},
  {Verkouter}, {Wells}, {Xiong}, {Szomoru}, {Griffith}, {Glendenning}, \&
  {Kern}}]{CASA_Team_etal_2022_SW_CASA}
{CASA Team}, {Bean}, B., {Bhatnagar}, S., {et~al.} 2022, \pasp, 134, 114501,
  \dodoi{10.1088/1538-3873/ac9642}

\bibitem[{{Castro Segura} {et~al.}(2020){Castro Segura}, {Altamirano},
  {Buisson}, {Degenaar}, {Diaz Trigo}, {Fender}, {Higginbottom}, {Knigge},
  {Long}, {Matthews}, {Mendez}, {Munoz Darias}, \&
  {Vincentelli}}]{Castro_Segura_etal_2020_HST_Proposal_16489C_Outflow_Legacy_Accretion_Survey}
{Castro Segura}, N., {Altamirano}, D., {Buisson}, D., {et~al.} 2020, {Outflow
  Legacy Accretion Survey: unveiling the wind driving mechanism in BHXRBs}, HST
  Proposal. Cycle 28, ID. \#16489

\bibitem[{{Castro-Tirado} {et~al.}(2023){Castro-Tirado}, {Sanchez-Ramirez},
  {Caballero-Garcia}, {Perez-Garcia}, {Fernandez-Garcia}, {Guziy}, {Hu},
  {Blazek}, {Hermelo}, {Pinter}, {Meintjes}, {van Heerden}, {Martin-Carrillo},
  {Hanlon}, {Hiriart}, {Lee}, {Carrasco-Garcia}, {Park}, {Gritsevich},
  {Castellon}, {Perez del Pulgar}, \&
  {Reina}}]{Castro-Tirado_etal_2023_ATel_16208_Swift_J1727_Optical_spectroscopy_confirms_BH-LMXB}
{Castro-Tirado}, A.~J., {Sanchez-Ramirez}, R., {Caballero-Garcia}, M.~D.,
  {et~al.} 2023, The Astronomer's Telegram, 16208, 1.
\newblock \url{https://www.astronomerstelegram.org/?read=16208}

\bibitem[{{Charles} {et~al.}(2019){Charles}, {Matthews}, {Buckley}, {Gandhi},
  {Kotze}, \&
  {Paice}}]{Charles_etal_2019_Swift_J1357.2-0933_HeII_outflows_2017_outburst}
{Charles}, P., {Matthews}, J.~H., {Buckley}, D. A.~H., {et~al.} 2019, \mnras,
  489, L47, \dodoi{10.1093/mnrasl/slz120}

\bibitem[{{Chatterjee} {et~al.}(2024){Chatterjee}, {Mondal}, {Singh}, \&
  {Sugizaki}}]{Chatterjee_etal_2024_Swift_J1727_Insight-HXMT_view_outburst}
{Chatterjee}, K., {Mondal}, S., {Singh}, C.~B., \& {Sugizaki}, M. 2024, \apj,
  977, 148, \dodoi{10.3847/1538-4357/ad8dc4}

\bibitem[{{Chauhan} {et~al.}(2019){Chauhan}, {Miller-Jones}, {Anderson},
  {Raja}, {Bahramian}, {Hotan}, {Indermuehle}, {Whiting}, {Allison},
  {Anderson}, {Bunton}, {Koribalski}, \&
  {Mahony}}]{Chauhan_etal_2019_MAXI_J1535-571_HI}
{Chauhan}, J., {Miller-Jones}, J.~C.~A., {Anderson}, G.~E., {et~al.} 2019,
  \mnras, 488, L129, \dodoi{10.1093/mnrasl/slz113}

\bibitem[{{Chauhan} {et~al.}(2021){Chauhan}, {Miller-Jones}, {Raja}, {Allison},
  {Jacob}, {Anderson}, {Carotenuto}, {Corbel}, {Fender}, {Hotan}, {Whiting},
  {Woudt}, {Koribalski}, \& {Mahony}}]{Chauhan_etal_2021_MAXI_J1348-630_HI}
{Chauhan}, J., {Miller-Jones}, J.~C.~A., {Raja}, W., {et~al.} 2021, \mnras,
  501, L60, \dodoi{10.1093/mnrasl/slaa195}

\bibitem[{{Chiang}(2023)}]{Chiang_2023_Corrected_SFD_CSFD_dust_map_minimal_extragalactic_contamination}
{Chiang}, Y.-K. 2023, \apj, 958, 118, \dodoi{10.3847/1538-4357/acf4a1}

\bibitem[{{Comrie} {et~al.}(2024){Comrie}, {Wang}, {Hwang}, {Moraghan},
  {Harris}, {Pińska}, {Raul-Omar}, {Chiang}, {Lin}, {Chang}, \&
  {Simmonds}}]{Comrie_etal_2024_SW_CARTA}
{Comrie}, A., {Wang}, K.-S., {Hwang}, Y.-H., {et~al.} 2024, CARTA: The Cube
  Analysis and Rendering Tool for Astronomy, 4.1.0,  Zenodo,
  \dodoi{10.5281/zenodo.15172686}

\bibitem[{{Dhawan} {et~al.}(2007){Dhawan}, {Mirabel}, {Rib{\'o}}, \&
  {Rodrigues}}]{Dhawan_etal_2007_GRS_1915_Kinematics_BH_XRB}
{Dhawan}, V., {Mirabel}, I.~F., {Rib{\'o}}, M., \& {Rodrigues}, I. 2007, \apj,
  668, 430, \dodoi{10.1086/520111}

\bibitem[{{Dickey}(1983)}]{Dickey_1983_Cygnus_X-3_HI_distance}
{Dickey}, J.~M. 1983, \apjl, 273, L71, \dodoi{10.1086/184132}

\bibitem[{{Draghis} {et~al.}(2023){Draghis}, {Miller}, {Homan}, {Uttley},
  {Bollemeijer}, {Steiner}, {Hare}, {Tombesi}, {Gendreau}, {Arzoumanian},
  {Strohmayer}, {Sanna}, {Altamirano}, {Buisson}, \&
  {Fabian}}]{Draghis_etal_2023_ATel_16219_Swift_J1727_prelim_specfit_QPO_evo_NICER_obs}
{Draghis}, P.~A., {Miller}, J.~M., {Homan}, J., {et~al.} 2023, The Astronomer's
  Telegram, 16219, 1

\bibitem[{{Dubus} {et~al.}(2001){Dubus}, {Kim}, {Menou}, {Szkody}, \&
  {Bowen}}]{Dubus_etal_2001_XTE_J1118+480_Optical_Spectroscopy_Outburst}
{Dubus}, G., {Kim}, R. S.~J., {Menou}, K., {Szkody}, P., \& {Bowen}, D.~V.
  2001, \apj, 553, 307, \dodoi{10.1086/320648}

\bibitem[{{Edenhofer} {et~al.}(2024){Edenhofer}, {Zucker}, {Frank}, {Saydjari},
  {Speagle}, {Finkbeiner}, \&
  {En{\ss}lin}}]{Edenhofer_etal_2024_pc_scale_3D_dust_map}
{Edenhofer}, G., {Zucker}, C., {Frank}, P., {et~al.} 2024, \aap, 685, A82,
  \dodoi{10.1051/0004-6361/202347628}

\bibitem[{{Fender} {et~al.}(2016){Fender}, {Woudt}, {Corbel}, {Coriat},
  {Daigne}, {Falcke}, {Girard}, {Heywood}, {Horesh}, {Horrell}, {Jonker},
  {Joseph}, {Kamble}, {Knigge}, {K{\"o}rding}, {Kotze}, {Kouveliotou}, {Lynch},
  {Maccarone}, {Meintjes}, {Migliari}, {Murphy}, {Nagayama}, {Nelemans},
  {Nicholson}, {O'Brien}, {Oodendaal}, {Oozeer}, {Osborne}, {P{\'e}rez-Torres},
  {Ratcliffe}, {Ribeiro}, {Rol}, {Rushton}, {Scaife}, {Schurch}, {Sivakoff},
  {Staley}, {Steeghs}, {Stewart}, {Swinbank}, {Vergani}, {Warner}, {Wiersema},
  {Armstrong}, {Groot}, {McBride}, {Miller-Jones}, {Mooley}, {Stappers},
  {Wijers}, {Bietenholz}, {Blyth}, {B{\"o}ttcher}, {Buckley}, {Charles},
  {Chomiuk}, {Coppejans}, {de Blok}, {van der Heyden}, {van der Horst}, \& {van
  Soelen}}]{Fender_etal_2016_ThunderKAT}
{Fender}, R., {Woudt}, P.~A., {Corbel}, S., {et~al.} 2016, in MeerKAT Science:
  On the Pathway to the SKA, 13, \dodoi{10.22323/1.277.0013}

\bibitem[{{Fitzpatrick}(2004)}]{Fitzpatrick_2004_Interstellar_Extinction_Milky_Way}
{Fitzpatrick}, E.~L. 2004, in Astronomical Society of the Pacific Conference
  Series, Vol. 309, Astrophysics of Dust, ed. A.~N. {Witt}, G.~C. {Clayton}, \&
  B.~T. {Draine}, 33, \dodoi{10.48550/arXiv.astro-ph/0401344}

\bibitem[{{Fragos} \&
  {McClintock}(2015)}]{Fragos+McClintock_2015_Origin_BH_Spin_LMXBs}
{Fragos}, T., \& {McClintock}, J.~E. 2015, \apj, 800, 17,
  \dodoi{10.1088/0004-637X/800/1/17}

\bibitem[{{Gaia Collaboration} {et~al.}(2016){Gaia Collaboration}, {Prusti},
  {de Bruijne}, {Brown}, {Vallenari}, {Babusiaux}, {Bailer-Jones}, {Bastian},
  {Biermann}, {Evans}, {Eyer}, {Jansen}, {Jordi}, {Klioner}, {Lammers},
  {Lindegren}, {Luri}, {Mignard}, {Milligan}, {Panem}, {Poinsignon},
  {Pourbaix}, {Randich}, {Sarri}, {Sartoretti}, {Siddiqui}, {Soubiran},
  {Valette}, {van Leeuwen}, {Walton}, {Aerts}, {Arenou}, {Cropper}, {Drimmel},
  {H{\o}g}, {Katz}, {Lattanzi}, {O'Mullane}, {Grebel}, {Holland}, {Huc},
  {Passot}, {Bramante}, {Cacciari}, {Casta{\~n}eda}, {Chaoul}, {Cheek}, {De
  Angeli}, {Fabricius}, {Guerra}, {Hern{\'a}ndez}, {Jean-Antoine-Piccolo},
  {Masana}, {Messineo}, {Mowlavi}, {Nienartowicz}, {Ord{\'o}{\~n}ez-Blanco},
  {Panuzzo}, {Portell}, {Richards}, {Riello}, {Seabroke}, {Tanga},
  {Th{\'e}venin}, {Torra}, {Els}, {Gracia-Abril}, {Comoretto},
  {Garcia-Reinaldos}, {Lock}, {Mercier}, {Altmann}, {Andrae}, {Astraatmadja},
  {Bellas-Velidis}, {Benson}, {Berthier}, {Blomme}, {Busso}, {Carry},
  {Cellino}, {Clementini}, {Cowell}, {Creevey}, {Cuypers}, {Davidson}, {De
  Ridder}, {de Torres}, {Delchambre}, {Dell'Oro}, {Ducourant}, {Fr{\'e}mat},
  {Garc{\'\i}a-Torres}, {Gosset}, {Halbwachs}, {Hambly}, {Harrison}, {Hauser},
  {Hestroffer}, {Hodgkin}, {Huckle}, {Hutton}, {Jasniewicz}, {Jordan},
  {Kontizas}, {Korn}, {Lanzafame}, {Manteiga}, {Moitinho}, {Muinonen},
  {Osinde}, {Pancino}, {Pauwels}, {Petit}, {Recio-Blanco}, {Robin}, {Sarro},
  {Siopis}, {Smith}, {Smith}, {Sozzetti}, {Thuillot}, {van Reeven}, {Viala},
  {Abbas}, {Abreu Aramburu}, {Accart}, {Aguado}, {Allan}, {Allasia},
  {Altavilla}, {{\'A}lvarez}, {Alves}, {Anderson}, {Andrei}, {Anglada Varela},
  {Antiche}, {Antoja}, {Ant{\'o}n}, {Arcay}, {Atzei}, {Ayache}, {Bach},
  {Baker}, {Balaguer-N{\'u}{\~n}ez}, {Barache}, {Barata}, {Barbier}, {Barblan},
  {Baroni}, {Barrado y Navascu{\'e}s}, {Barros}, {Barstow}, {Becciani},
  {Bellazzini}, {Bellei}, {Bello Garc{\'\i}a}, {Belokurov}, {Bendjoya},
  {Berihuete}, {Bianchi}, {Bienaym{\'e}}, {Billebaud}, {Blagorodnova},
  {Blanco-Cuaresma}, {Boch}, {Bombrun}, {Borrachero}, {Bouquillon}, {Bourda},
  {Bouy}, {Bragaglia}, {Breddels}, {Brouillet}, {Br{\"u}semeister},
  {Bucciarelli}, {Budnik}, {Burgess}, {Burgon}, {Burlacu}, {Busonero}, {Buzzi},
  {Caffau}, {Cambras}, {Campbell}, {Cancelliere}, {Cantat-Gaudin}, {Carlucci},
  {Carrasco}, {Castellani}, {Charlot}, {Charnas}, {Charvet}, {Chassat},
  {Chiavassa}, {Clotet}, {Cocozza}, {Collins}, {Collins}, \&
  {Costigan}}]{Gaia_Collaboration_etal_2016_The_Gaia_Mission}
{Gaia Collaboration}, {Prusti}, T., {de Bruijne}, J.~H.~J., {et~al.} 2016,
  \aap, 595, A1, \dodoi{10.1051/0004-6361/201629272}

\bibitem[{{Gandhi} {et~al.}(2019){Gandhi}, {Rao}, {Johnson}, {Paice}, \&
  {Maccarone}}]{Gandhi_etal_2019_Gaia_DR2_distances_peculiar_velocities_for_GBHTs}
{Gandhi}, P., {Rao}, A., {Johnson}, M. A.~C., {Paice}, J.~A., \& {Maccarone},
  T.~J. 2019, \mnras, 485, 2642, \dodoi{10.1093/mnras/stz438}

\bibitem[{{Green}(2018)}]{Green_2018_SW_dustmaps}
{Green}, G. 2018, The Journal of Open Source Software, 3, 695,
  \dodoi{10.21105/joss.00695}

\bibitem[{{Green} {et~al.}(2019){Green}, {Schlafly}, {Zucker}, {Speagle}, \&
  {Finkbeiner}}]{Green_etal_2019_3D_dust_map_Gaia_Pan-STARRS1_2MASS}
{Green}, G.~M., {Schlafly}, E., {Zucker}, C., {Speagle}, J.~S., \&
  {Finkbeiner}, D. 2019, \apj, 887, 93, \dodoi{10.3847/1538-4357/ab5362}

\bibitem[{{Gualandris} {et~al.}(2005){Gualandris}, {Colpi}, {Portegies Zwart},
  \& {Possenti}}]{Gualandris_etal_2005_XTE_J1118+480_Asymmetric_Natal_Kick}
{Gualandris}, A., {Colpi}, M., {Portegies Zwart}, S., \& {Possenti}, A. 2005,
  \apj, 618, 845, \dodoi{10.1086/426126}

\bibitem[{{G{\"u}ver} \&
  {{\"O}zel}(2009)}]{Guver_etal_2009_Optical_extinction_hydrogen_column_density}
{G{\"u}ver}, T., \& {{\"O}zel}, F. 2009, \mnras, 400, 2050,
  \dodoi{10.1111/j.1365-2966.2009.15598.x}

\bibitem[{Hack {et~al.}(2018)Hack, Dencheva, Sontag, Sosey, \&
  Droettboom}]{Hack_etal_2018_SW_stistools}
Hack, W., Dencheva, N., Sontag, C., Sosey, M., \& Droettboom, M. 2018, STIS
  Python User Tools.
\newblock \url{https://stistools.readthedocs.io/en/latest/index.html}

\bibitem[{Harris {et~al.}(2020)Harris, Millman, van~der Walt, Gommers,
  Virtanen, Cournapeau, Wieser, Taylor, Berg, Smith, Kern, Picus, Hoyer, van
  Kerkwijk, Brett, Haldane, del R{\'{i}}o, Wiebe, Peterson,
  G{\'{e}}rard-Marchant, Sheppard, Reddy, Weckesser, Abbasi, Gohlke, \&
  Oliphant}]{Harris_etal_2020_SW_NumPy}
Harris, C.~R., Millman, K.~J., van~der Walt, S.~J., {et~al.} 2020, Nature, 585,
  357, \dodoi{10.1038/s41586-020-2649-2}

\bibitem[{{Heinz} {et~al.}(2015){Heinz}, {Burton}, {Braiding}, {Brandt},
  {Jonker}, {Sell}, {Fender}, {Nowak}, \&
  {Schulz}}]{Heinz_etal_2015_Cir_X-1_LotR_Kinematic_Distance_X-ray_Light_Echo}
{Heinz}, S., {Burton}, M., {Braiding}, C., {et~al.} 2015, \apj, 806, 265,
  \dodoi{10.1088/0004-637X/806/2/265}

\bibitem[{{HI4PI Collaboration} {et~al.}(2016){HI4PI Collaboration}, {Ben
  Bekhti}, {Fl{\"o}er}, {Keller}, {Kerp}, {Lenz}, {Winkel}, {Bailin},
  {Calabretta}, {Dedes}, {Ford}, {Gibson}, {Haud}, {Janowiecki}, {Kalberla},
  {Lockman}, {McClure-Griffiths}, {Murphy}, {Nakanishi}, {Pisano}, \&
  {Staveley-Smith}}]{HI4PI_Collaboration_etal_2016_Fullsky_H_I_survey_EBHIS_GASS}
{HI4PI Collaboration}, {Ben Bekhti}, N., {Fl{\"o}er}, L., {et~al.} 2016, \aap,
  594, A116, \dodoi{10.1051/0004-6361/201629178}

\bibitem[{{Hunter} {et~al.}(2024){Hunter}, {Sormani}, {Beckmann}, {Vasiliev},
  {Glover}, {Klessen}, {Soler}, {Brucy}, {Girichidis}, {G{\"o}ller}, {Ohlin},
  {Tress}, {Molinari}, {Gerhard}, {Benedettini}, {Smith}, {Hennebelle}, \&
  {Testi}}]{Hunter_etal_2024_Testing_Galactic_kinematic_distances}
{Hunter}, G.~H., {Sormani}, M.~C., {Beckmann}, J.~P., {et~al.} 2024, \aap, 692,
  A216, \dodoi{10.1051/0004-6361/202450000}

\bibitem[{Hunter(2007)}]{Hunter_2007_SW_Matplotlib}
Hunter, J.~D. 2007, Computing in Science \& Engineering, 9, 90,
  \dodoi{10.1109/MCSE.2007.55}

\bibitem[{{Hynes} {et~al.}(2002){Hynes}, {Haswell}, {Chaty}, {Shrader}, \&
  {Cui}}]{Hynes_etal_2002_XTE_J1859_Accretion_Disc_X-ray_Distance}
{Hynes}, R.~I., {Haswell}, C.~A., {Chaty}, S., {Shrader}, C.~R., \& {Cui}, W.
  2002, \mnras, 331, 169, \dodoi{10.1046/j.1365-8711.2002.05175.x}

\bibitem[{Icazatti {et~al.}(2023)Icazatti, Abril-Pla, Klami, \&
  Martin}]{Icazatti_etal_2023_SW_PreliZ}
Icazatti, A., Abril-Pla, O., Klami, A., \& Martin, O.~A. 2023, Journal of Open
  Source Software, 8, 5499, \dodoi{10.21105/joss.05499}

\bibitem[{{Jonker} \&
  {Nelemans}(2004)}]{Jonker+Nelemans_2004_Distances_Galactic_LMXB_spectroscopy}
{Jonker}, P.~G., \& {Nelemans}, G. 2004, \mnras, 354, 355,
  \dodoi{10.1111/j.1365-2966.2004.08193.x}

\bibitem[{{Kalberla} \& {Kerp}(2009)}]{Kalberla_Kerp_2009_MW_HI}
{Kalberla}, P. M.~W., \& {Kerp}, J. 2009, \araa, 47, 27,
  \dodoi{10.1146/annurev-astro-082708-101823}

\bibitem[{{Kalemci} {et~al.}(2013){Kalemci}, {Din{\c{c}}er}, {Tomsick},
  {Buxton}, {Bailyn}, \&
  {Chun}}]{Kalemci_etal_2013_BHT_distance_state_transitions}
{Kalemci}, E., {Din{\c{c}}er}, T., {Tomsick}, J.~A., {et~al.} 2013, \apj, 779,
  95, \dodoi{10.1088/0004-637X/779/2/95}

\bibitem[{{Kluyver} {et~al.}(2016){Kluyver}, {Ragan-Kelley}, {P{\'e}rez},
  {Granger}, {Bussonnier}, {Frederic}, {Kelley}, {Hamrick}, {Grout}, {Corlay},
  {Ivanov}, {Avila}, {Abdalla}, {Willing}, \& {Jupyter Development
  Team}}]{Kluyver_etal_2016_SW_Jupyter}
{Kluyver}, T., {Ragan-Kelley}, B., {P{\'e}rez}, F., {et~al.} 2016, in IOS
  Press, 87--90, \dodoi{10.3233/978-1-61499-649-1-87}

\bibitem[{Kumar {et~al.}(2019)Kumar, Carroll, Hartikainen, \&
  Martin}]{Kumar_etal_2019_SW_ArviZ}
Kumar, R., Carroll, C., Hartikainen, A., \& Martin, O. 2019, Journal of Open
  Source Software, 4, 1143, \dodoi{10.21105/joss.01143}

\bibitem[{{Lamer} {et~al.}(2021){Lamer}, {Schwope}, {Predehl}, {Traulsen},
  {Wilms}, \&
  {Freyberg}}]{Lamer_etal_2021_X-ray_scattering_distance_MAXI_J1348}
{Lamer}, G., {Schwope}, A.~D., {Predehl}, P., {et~al.} 2021, \aap, 647, A7,
  \dodoi{10.1051/0004-6361/202039757}

\bibitem[{{Liu} {et~al.}(2024){Liu}, {Xu}, {Zhang}, {Yu}, {Huang}, {Tao},
  {Zhang}, {Yang}, {Zhao}, {Qu}, \&
  {Song}}]{Liu_etal_2024_Swift_J1727_Broadband_X-ray_Spectral_Properties_Rising_Outburst}
{Liu}, H.-X., {Xu}, Y.-J., {Zhang}, S.-N., {et~al.} 2024, arXiv e-prints,
  arXiv:2406.03834, \dodoi{10.48550/arXiv.2406.03834}

\bibitem[{{Lockman} {et~al.}(2007){Lockman}, {Blundell}, \&
  {Goss}}]{Lockman_etal_2007_SS433_HI_distance_ISM}
{Lockman}, F.~J., {Blundell}, K.~M., \& {Goss}, W.~M. 2007, \mnras, 381, 881,
  \dodoi{10.1111/j.1365-2966.2007.12170.x}

\bibitem[{{Maccarone}(2002)}]{Maccarone_2002_Microquasar_jet_misalignment}
{Maccarone}, T.~J. 2002, \mnras, 336, 1371,
  \dodoi{10.1046/j.1365-8711.2002.05876.x}

\bibitem[{{Maccarone} {et~al.}(2020){Maccarone}, {Osler}, {Miller-Jones},
  {Atri}, {Russell}, {Meier}, {McHardy}, \&
  {Longa-Pe{\~n}a}}]{Maccarone_etal_2020_4U_1957+11_Upper_limit_jet_power}
{Maccarone}, T.~J., {Osler}, A., {Miller-Jones}, J. C.~A., {et~al.} 2020,
  \mnras, 498, L40, \dodoi{10.1093/mnrasl/slaa120}

\bibitem[{{Martin} {et~al.}(2008){Martin}, {Tout}, \&
  {Pringle}}]{Martin_etal_2008_GRO_J1655-40_Alignment_timescale_microquasar}
{Martin}, R.~G., {Tout}, C.~A., \& {Pringle}, J.~E. 2008, \mnras, 387, 188,
  \dodoi{10.1111/j.1365-2966.2008.13148.x}

\bibitem[{{Mata S{\'a}nchez} {et~al.}(2024){Mata S{\'a}nchez},
  {Mu{\~n}oz-Darias}, {Armas Padilla}, {Casares}, \&
  {Torres}}]{MataSanchez_etal_2024_Swift_J1727_Inflows_Outflows}
{Mata S{\'a}nchez}, D., {Mu{\~n}oz-Darias}, T., {Armas Padilla}, M., {Casares},
  J., \& {Torres}, M.~A.~P. 2024, \aap, 682, L1,
  \dodoi{10.1051/0004-6361/202348754}

\bibitem[{{Mata S{\'a}nchez} {et~al.}(2025){Mata S{\'a}nchez}, {Torres},
  {Casares}, {Mu{\~n}oz-Darias}, {Armas Padilla}, \&
  {Yanes-Rizo}}]{MataSanchez_etal_2025_Swift_J1727_Dynamically_Confirmed_BH}
{Mata S{\'a}nchez}, D., {Torres}, M.~A.~P., {Casares}, J., {et~al.} 2025, \aap,
  693, A129, \dodoi{10.1051/0004-6361/202451960}

\bibitem[{{Megier} {et~al.}(2009){Megier}, {Strobel}, {Galazutdinov}, \&
  {Kre{\l}owski}}]{Megier_etal_2009_Interstellar_CaII_and_Distance}
{Megier}, A., {Strobel}, A., {Galazutdinov}, G.~A., \& {Kre{\l}owski}, J. 2009,
  \aap, 507, 833, \dodoi{10.1051/0004-6361/20079144}

\bibitem[{{Miller-Jones} {et~al.}(2023{\natexlab{a}}){Miller-Jones},
  {Bahramian}, {Altamirano}, {Homan}, {Russell}, \&
  {Sivakoff}}]{Miller-Jones_etal_2023_ATel_16271_Swift_J1727_Radio_quenching_subsequent_flaring}
{Miller-Jones}, J.~C.~A., {Bahramian}, A., {Altamirano}, D., {et~al.}
  2023{\natexlab{a}}, The Astronomer's Telegram, 16271, 1.
\newblock \url{https://www.astronomerstelegram.org/?read=16271}

\bibitem[{{Miller-Jones} {et~al.}(2009){Miller-Jones}, {Jonker}, {Dhawan},
  {Brisken}, {Rupen}, {Nelemans}, \&
  {Gallo}}]{Miller-Jones_etal_2009_V404_Cyg_parallax}
{Miller-Jones}, J.~C.~A., {Jonker}, P.~G., {Dhawan}, V., {et~al.} 2009, \apjl,
  706, L230, \dodoi{10.1088/0004-637X/706/2/L230}

\bibitem[{{Miller-Jones} {et~al.}(2023{\natexlab{b}}){Miller-Jones},
  {Sivakoff}, {Bahramian}, \&
  {Russell}}]{Miller-Jones_etal_2023_ATel_16211_Swift_J1727_VLA_radio_detection}
{Miller-Jones}, J.~C.~A., {Sivakoff}, G.~R., {Bahramian}, A., \& {Russell},
  T.~D. 2023{\natexlab{b}}, The Astronomer's Telegram, 16211, 1.
\newblock \url{https://www.astronomerstelegram.org/?read=16211}

\bibitem[{{Miller-Jones} {et~al.}(2021){Miller-Jones}, {Bahramian}, {Orosz},
  {Mandel}, {Gou}, {Maccarone}, {Neijssel}, {Zhao}, {Zi{\'o}{\l}kowski},
  {Reid}, {Uttley}, {Zheng}, {Byun}, {Dodson}, {Grinberg}, {Jung}, {Kim},
  {Marcote}, {Markoff}, {Rioja}, {Rushton}, {Russell}, {Sivakoff}, {Tetarenko},
  {Tudose}, \& {Wilms}}]{Miller-Jones_etal_2021_Cyg_X-1_parallax_revised}
{Miller-Jones}, J. C.~A., {Bahramian}, A., {Orosz}, J.~A., {et~al.} 2021,
  Science, 371, 1046, \dodoi{10.1126/science.abb3363}

\bibitem[{{Mirabel} \&
  {Rodr{\'\i}guez}(1994)}]{Mirabel+Rodriguez_1994_GRS_1915_Nature_Superluminal}
{Mirabel}, I.~F., \& {Rodr{\'\i}guez}, L.~F. 1994, \nat, 371, 46,
  \dodoi{10.1038/371046a0}

\bibitem[{{Munari} \& {Zwitter}(1997)}]{Munari_Zwitter_1997_EWs_NaI_KI_EBmV}
{Munari}, U., \& {Zwitter}, T. 1997, \aap, 318, 269

\bibitem[{{Negoro} {et~al.}(2023){Negoro}, {Serino}, {Nakajima}, {Kobayashi},
  {Tanaka}, {Soejima}, {Kudo}, {Mihara}, {Kawamuro}, {Yamada}, {Tamagawa},
  {Kawai}, {Matsuoka}, {Sakamoto}, {Sugita}, {Hiramatsu}, {Nishikawa},
  {Yoshida}, {Tsuboi}, {Urabe}, {Nawa}, {Nemoto}, {Shidatsu}, {Takahashi},
  {Niwano}, {Sato}, {Higuchi}, {Yatsu}, {Nakahira}, {Ueno}, {Tomida},
  {Ishikawa}, {Ogawa}, {Kurihara}, {Ueda}, {Setoguchi}, {Yoshitake},
  {Nakatani}, {Yamauchi}, {Hagiwara}, {Umeki}, {Otsuki}, {Yamaoka}, {Kawakubo},
  {Sugizaki}, \&
  {Iwakiri}}]{Negoro_etal_2023_ATel_16205_Swift_J1727_MAXI/GSC_detection}
{Negoro}, H., {Serino}, M., {Nakajima}, M., {et~al.} 2023, The Astronomer's
  Telegram, 16205, 1.
\newblock \url{https://www.astronomerstelegram.org/?read=16205}

\bibitem[{{O'Connor} {et~al.}(2023){O'Connor}, {Hare}, {Younes}, {Gendreau},
  {Arzoumanian}, \&
  {Ferrara}}]{O'Connor_etal_2023_ATel_16207_Swift_J1727_NICER_detection}
{O'Connor}, B., {Hare}, J., {Younes}, G., {et~al.} 2023, The Astronomer's
  Telegram, 16207, 1

\bibitem[{{Paczy{\'n}ski}(1967)}]{Paczynski_1967_Evolution_Close_Binaries_Wolf-Rayet_Stars}
{Paczy{\'n}ski}, B. 1967, \actaa, 17, 355

\bibitem[{{Peng} {et~al.}(2024){Peng}, {Zhang}, {Shui}, {Zhang}, {Kong},
  {Chen}, {Wang}, {Ji}, {Qu}, {Tao}, {Ge}, {Chang}, {Li}, {Li}, {Yu}, \&
  {Yan}}]{Peng_etal_2024_Swift_J1727_NICER_NuSTAR_Insight-HXMT_views}
{Peng}, J.-Q., {Zhang}, S., {Shui}, Q.-C., {et~al.} 2024, \apjl, 960, L17,
  \dodoi{10.3847/2041-8213/ad17ca}

\bibitem[{{Podsiadlowski} {et~al.}(2003){Podsiadlowski}, {Rappaport}, \&
  {Han}}]{Podsiadlowski+Rappaport+Han_2003_Formation_Evolution_BH_Binaries}
{Podsiadlowski}, P., {Rappaport}, S., \& {Han}, Z. 2003, \mnras, 341, 385,
  \dodoi{10.1046/j.1365-8711.2003.06464.x}

\bibitem[{{Powell} {et~al.}(2007){Powell}, {Haswell}, \&
  {Falanga}}]{Powell_etal_2007_Mass_Transfer_LMX_Transient_Decays_Distance}
{Powell}, C.~R., {Haswell}, C.~A., \& {Falanga}, M. 2007, \mnras, 374, 466,
  \dodoi{10.1111/j.1365-2966.2006.11144.x}

\bibitem[{{Reid}(2022)}]{Reid_2022_On_the_accuracy_of_3D_kinematic_distances}
{Reid}, M.~J. 2022, \aj, 164, 133, \dodoi{10.3847/1538-3881/ac80bb}

\bibitem[{{Reid} {et~al.}(2011){Reid}, {McClintock}, {Narayan}, {Gou},
  {Remillard}, \& {Orosz}}]{Reid_etal_2011_Cyg_X-1_Trig_VLBI_parallax}
{Reid}, M.~J., {McClintock}, J.~E., {Narayan}, R., {et~al.} 2011, \apj, 742,
  83, \dodoi{10.1088/0004-637X/742/2/83}

\bibitem[{{Reid} {et~al.}(2014{\natexlab{a}}){Reid}, {McClintock}, {Steiner},
  {Steeghs}, {Remillard}, {Dhawan}, \&
  {Narayan}}]{Reid_etal_2014_GRS_1915_Parallax_Distance_Revised_BH_Mass}
{Reid}, M.~J., {McClintock}, J.~E., {Steiner}, J.~F., {et~al.}
  2014{\natexlab{a}}, \apj, 796, 2, \dodoi{10.1088/0004-637X/796/1/2}

\bibitem[{{Reid} \&
  {Miller-Jones}(2023)}]{Reid+Miller-Jones_2023_Cyg_X-3_GRS_1915_distances_parallax}
{Reid}, M.~J., \& {Miller-Jones}, J.~C.~A. 2023, \apj, 959, 85,
  \dodoi{10.3847/1538-4357/acfe0c}

\bibitem[{{Reid} {et~al.}(2014{\natexlab{b}}){Reid}, {Menten}, {Brunthaler},
  {Zheng}, {Dame}, {Xu}, {Wu}, {Zhang}, {Sanna}, {Sato}, {Hachisuka}, {Choi},
  {Immer}, {Moscadelli}, {Rygl}, \&
  {Bartkiewicz}}]{Reid_etal_2014_Trig_Parallax_HMSFRs_Structure_Kinematics_MW}
{Reid}, M.~J., {Menten}, K.~M., {Brunthaler}, A., {et~al.} 2014{\natexlab{b}},
  \apj, 783, 130, \dodoi{10.1088/0004-637X/783/2/130}

\bibitem[{{Reid} {et~al.}(2019){Reid}, {Menten}, {Brunthaler}, {Zheng}, {Dame},
  {Xu}, {Li}, {Sakai}, {Wu}, {Immer}, {Zhang}, {Sanna}, {Moscadelli}, {Rygl},
  {Bartkiewicz}, {Hu}, {Quiroga-Nu{\~n}ez}, \& {van
  Langevelde}}]{Reid_etal_2019_Trig_Parallax_HMSFRs_MW_Our_View}
---. 2019, \apj, 885, 131, \dodoi{10.3847/1538-4357/ab4a11}

\bibitem[{{Rybarczyk} {et~al.}(2024){Rybarczyk}, {Wenger}, \&
  {Stanimirovi{\'c}}}]{Rybarczyk_Wenger_Stanimirovic_2024_Vert_Dist_HI_Clouds_21cm_Abs_High_Gal_Lat}
{Rybarczyk}, D.~R., {Wenger}, T.~V., \& {Stanimirovi{\'c}}, S. 2024, \apj, 975,
  167, \dodoi{10.3847/1538-4357/ad79f7}

\bibitem[{{Sault} {et~al.}(1995){Sault}, {Teuben}, \&
  {Wright}}]{Sault_etal_1995_SW_Miriad}
{Sault}, R.~J., {Teuben}, P.~J., \& {Wright}, M.~C.~H. 1995, in Astronomical
  Society of the Pacific Conference Series, Vol.~77, Astronomical Data Analysis
  Software and Systems IV, ed. R.~A. {Shaw}, H.~E. {Payne}, \& J.~J.~E.
  {Hayes}, 433, \dodoi{10.48550/arXiv.astro-ph/0612759}

\bibitem[{{Savage} \&
  {Mathis}(1979)}]{Savage_Mathis_1979_Properties_Interstellar_Dust}
{Savage}, B.~D., \& {Mathis}, J.~S. 1979, \araa, 17, 73,
  \dodoi{10.1146/annurev.aa.17.090179.000445}

\bibitem[{{Schlafly} \&
  {Finkbeiner}(2011)}]{Schlafly_Finkbeiner_2011_EBmV_with_SDSS}
{Schlafly}, E.~F., \& {Finkbeiner}, D.~P. 2011, \apj, 737, 103,
  \dodoi{10.1088/0004-637X/737/2/103}

\bibitem[{{Schlegel} {et~al.}(1998){Schlegel}, {Finkbeiner}, \&
  {Davis}}]{Schlegel_Finkbeiner_David_1998_SFD_dust_map_via_IR_emission_for_reddening_and_CMBRFs}
{Schlegel}, D.~J., {Finkbeiner}, D.~P., \& {Davis}, M. 1998, \apj, 500, 525,
  \dodoi{10.1086/305772}

\bibitem[{{Svoboda} {et~al.}(2024){Svoboda}, {Dov{\v{c}}iak}, {Steiner},
  {Kaaret}, {Podgorn{\'y}}, {Poutanen}, {Veledina}, {Muleri}, {Taverna},
  {Krawczynski}, {Brigitte}, {Datta}, {Bianchi}, {Mu{\~n}oz-Darias}, {Negro},
  {Rodriguez Cavero}, {Castro Segura}, {Bollemeijer}, {Garc{\'\i}a}, {Ingram},
  {Matt}, {Nathan}, {Weisskopf}, {Altamirano}, {Baldini}, {Capitanio}, {Egron},
  {Emami}, {Hu}, {Marra}, {Mastroserio}, {Petrucci}, {Ratheesh}, {Soffitta},
  {Tombesi}, {Yang}, \&
  {Zhang}}]{Svoboda_etal_2024_Swift_J1727_X-ray_pol_drop_in_soft_state}
{Svoboda}, J., {Dov{\v{c}}iak}, M., {Steiner}, J.~F., {et~al.} 2024, \apjl,
  966, L35, \dodoi{10.3847/2041-8213/ad402e}

\bibitem[{{Tetarenko} {et~al.}(2016){Tetarenko}, {Sivakoff}, {Heinke}, \&
  {Gladstone}}]{Tetarenko_etal_2016_WATCHDOG}
{Tetarenko}, B.~E., {Sivakoff}, G.~R., {Heinke}, C.~O., \& {Gladstone}, J.~C.
  2016, \apjs, 222, 15, \dodoi{10.3847/0067-0049/222/2/15}

\bibitem[{{Vahdat Motlagh} {et~al.}(2019){Vahdat Motlagh}, {Kalemci}, \&
  {Maccarone}}]{VahdatMotlagh+Kalemci+Maccarone_2019_State_transition_luminosities_Galactic_BHT_outburst_decay}
{Vahdat Motlagh}, A., {Kalemci}, E., \& {Maccarone}, T.~J. 2019, \mnras, 485,
  2744, \dodoi{10.1093/mnras/stz569}

\bibitem[{{Veledina} {et~al.}(2023){Veledina}, {Muleri}, {Dov{\v{c}}iak},
  {Poutanen}, {Ratheesh}, {Capitanio}, {Matt}, {Soffitta}, {Tennant}, {Negro},
  {Kaaret}, {Costa}, {Ingram}, {Svoboda}, {Krawczynski}, {Bianchi}, {Steiner},
  {Garc{\'\i}a}, {Kravtsov}, {Nitindala}, {Ewing}, {Mastroserio}, {Marinucci},
  {Ursini}, {Tombesi}, {Tsygankov}, {Yang}, {Weisskopf}, {Trushkin}, {Egron},
  {Iacolina}, {Pilia}, {Marra}, {Miku{\v{s}}incov{\'a}}, {Nathan}, {Parra},
  {Petrucci}, {Podgorn{\'y}}, {Tugliani}, {Zane}, {Zhang}, {Agudo},
  {Antonelli}, {Bachetti}, {Baldini}, {Baumgartner}, {Bellazzini}, {Bongiorno},
  {Bonino}, {Brez}, {Bucciantini}, {Castellano}, {Cavazzuti}, {Chen},
  {Ciprini}, {De Rosa}, {Del Monte}, {Di Gesu}, {Di Lalla}, {Di Marco},
  {Donnarumma}, {Doroshenko}, {Ehlert}, {Enoto}, {Evangelista}, {Fabiani},
  {Ferrazzoli}, {Gunji}, {Hayashida}, {Heyl}, {Iwakiri}, {Jorstad}, {Karas},
  {Kislat}, {Kitaguchi}, {Kolodziejczak}, {La Monaca}, {Latronico}, {Liodakis},
  {Maldera}, {Manfreda}, {Marin}, {Marscher}, {Marshall}, {Massaro},
  {Mitsuishi}, {Mizuno}, {Ng}, {O'Dell}, {Omodei}, {Oppedisano}, {Papitto},
  {Pavlov}, {Peirson}, {Perri}, {Pesce-Rollins}, {Possenti}, {Puccetti},
  {Ramsey}, {Rankin}, {Roberts}, {Romani}, {Sgr{\`o}}, {Slane}, {Spandre},
  {Swartz}, {Tamagawa}, {Tavecchio}, {Taverna}, {Tawara}, {Thomas}, {Trois},
  {Turolla}, {Vink}, {Wu}, \&
  {Xie}}]{Veledina_etal_2023_Swift_J1727_X-ray_pol_flux_scaling_distance}
{Veledina}, A., {Muleri}, F., {Dov{\v{c}}iak}, M., {et~al.} 2023, \apjl, 958,
  L16, \dodoi{10.3847/2041-8213/ad0781}

\bibitem[{Virtanen {et~al.}(2020)Virtanen, Gommers, Oliphant, Haberland, Reddy,
  Cournapeau, Burovski, Peterson, Weckesser, Bright, {van der Walt}, Brett,
  Wilson, Millman, Mayorov, Nelson, Jones, Kern, Larson, Carey, Polat, Feng,
  Moore, {VanderPlas}, Laxalde, Perktold, Cimrman, Henriksen, Quintero, Harris,
  Archibald, Ribeiro, Pedregosa, {van Mulbregt}, \& {SciPy 1.0
  Contributors}}]{Pauli_etal_2020_SW_SciPy}
Virtanen, P., Gommers, R., Oliphant, T.~E., {et~al.} 2020, Nature Methods, 17,
  261, \dodoi{10.1038/s41592-019-0686-2}

\bibitem[{{Wallerstein} {et~al.}(2007){Wallerstein}, {Sandstrom}, \&
  {Gredel}}]{Wallerstein_Sandstrom_Gredel_2007_8621_Angstroms_EBmV}
{Wallerstein}, G., {Sandstrom}, K., \& {Gredel}, R. 2007, \pasp, 119, 1268,
  \dodoi{10.1086/521835}

\bibitem[{Wenger(2018)}]{Wenger_2018_SW_KDCT_Zenodo_DOI_1166001}
Wenger, T.~V. 2018, tvwenger/kd v1.0, v1.0,  Zenodo,
  \dodoi{10.5281/zenodo.1166001}

\bibitem[{{Wenger} {et~al.}(2018){Wenger}, {Balser}, {Anderson}, \&
  {Bania}}]{Wenger_etal_2018_Kinematic_Distances_MC}
{Wenger}, T.~V., {Balser}, D.~S., {Anderson}, L.~D., \& {Bania}, T.~M. 2018,
  \apj, 856, 52, \dodoi{10.3847/1538-4357/aaaec8}

\bibitem[{{Wilms} {et~al.}(2000){Wilms}, {Allen}, \&
  {McCray}}]{Wilms_etal_2000_ISM_X-ray_Absorption}
{Wilms}, J., {Allen}, A., \& {McCray}, R. 2000, \apj, 542, 914,
  \dodoi{10.1086/317016}

\bibitem[{{Wood} {et~al.}(2024){Wood}, {Miller-Jones}, {Bahramian}, {Tingay},
  {Prabu}, {Russell}, {Atri}, {Carotenuto}, {Altamirano}, {Motta}, {Hyland},
  {Reynolds}, {Weston}, {Fender}, {K{\"o}rding}, {Maitra}, {Markoff},
  {Migliari}, {Russell}, {Sarazin}, {Sivakoff}, {Soria}, {Tetarenko}, \&
  {Tudose}}]{Wood_etal_2024_Swift_J1727_Largest_Resolved_Jet_XRB}
{Wood}, C.~M., {Miller-Jones}, J. C.~A., {Bahramian}, A., {et~al.} 2024, \apjl,
  971, L9, \dodoi{10.3847/2041-8213/ad6572}

\bibitem[{{Wood} {et~al.}(2025){Wood}, {Miller-Jones}, {Bahramian}, {Tingay},
  {Liu}, {Altamirano}, {Fender}, {K{\"o}rding}, {Maitra}, {Markoff}, {Russell},
  {Russell}, {Sarazin}, {Sivakoff}, {Soria}, {Tetarenko}, \&
  {Tudose}}]{Wood_etal_2025_Swift_J1727_Ejection_Transient_Jets_Time_Dependent_Visibility_Modelling}
---. 2025, arXiv e-prints, arXiv:2503.03073, \dodoi{10.48550/arXiv.2503.03073}

\bibitem[{{Woodgate} {et~al.}(1998){Woodgate}, {Kimble}, {Bowers}, {Kraemer},
  {Kaiser}, {Danks}, {Grady}, {Loiacono}, {Brumfield}, {Feinberg}, {Gull},
  {Heap}, {Maran}, {Lindler}, {Hood}, {Meyer}, {Vanhouten}, {Argabright},
  {Franka}, {Bybee}, {Dorn}, {Bottema}, {Woodruff}, {Michika}, {Sullivan},
  {Hetlinger}, {Ludtke}, {Stocker}, {Delamere}, {Rose}, {Becker}, {Garner},
  {Timothy}, {Blouke}, {Joseph}, {Hartig}, {Green}, {Jenkins}, {Linsky},
  {Hutchings}, {Moos}, {Boggess}, {Roesler}, \&
  {Weistrop}}]{Woodgate_etal_1998_HST_STIS_Design}
{Woodgate}, B.~E., {Kimble}, R.~A., {Bowers}, C.~W., {et~al.} 1998, \pasp, 110,
  1183, \dodoi{10.1086/316243}

\bibitem[{{Zdziarski} {et~al.}(2025){Zdziarski}, {Wood}, \&
  {Carotenuto}}]{Zdziarski+Wood+Carotenuto_2025_Swift_J1727_Modeling_Extended_Emission_Compact_Jets}
{Zdziarski}, A.~A., {Wood}, C.~M., \& {Carotenuto}, F. 2025, arXiv e-prints,
  arXiv:2504.20962, \dodoi{10.48550/arXiv.2504.20962}

\bibitem[{{Zhu} {et~al.}(2017){Zhu}, {Tian}, {Li}, \&
  {Zhang}}]{Zhu_etal_2017_Galactic_gas_to_extinction_and_distribution}
{Zhu}, H., {Tian}, W., {Li}, A., \& {Zhang}, M. 2017, \mnras, 471, 3494,
  \dodoi{10.1093/mnras/stx1580}

\end{thebibliography}
\bibliographystyle{aasjournal}

\end{document}